\documentclass[11pt,twoside]{article}
\pdfoutput=1
\usepackage{jcst}
\usepackage[utf8]{inputenc}
\usepackage{cite}
\usepackage{setspace}
\usepackage{amssymb}
\usepackage{amsmath}
\usepackage{multirow}
\usepackage{subfig}
\usepackage{color}
\usepackage{multicol}
\usepackage{algorithm}
\usepackage{algpseudocode}
\newtheorem{mydef}{Definition}

\begin{document}
\begin{singlespacing}

\setcounter{page}{1}
\setpageinformation
{Ouriques JFS, Cartaxo EG, Alves ELG, Machado PDL}
{}{}{}{Mar.}{2017}
{Ouriques {\it et al}.: A Hint-Based for MBT Test Case Prioritization}

\title{A Hint-Based Technique for System Level Model-Based Test Case Prioritization}

\author{Jo\~{a}o F. S. Ouriques$^1$, Emanuela G. Cartaxo$^1$, Everton L. G. Alves$^1$, and Patr\'{i}cia D. L. Machado$^1$}

\address{$^1$Software Practices Laboratory, Federal University of Campina Grande, Para\'{i}ba, 58429-900, Brazil
}

\email{\{jfelipe, emanuela\}@copin.ufcg.edu.br, \{everton, patricia\}@computacao.ufcg.edu.br}
\received{March 6, 2017}

\footnotetext{This work was partially supported by the National Institute of Science and Technology for Software Engineering (2015), funded by CNPq/Brasil, grant 573964/2008-4. First author was also supported by CNPq grant 141215/2012-7.}

\begin{abstract}
Test Case Prioritization (TCP) techniques aim at proposing new test case execution orders to favor the achievement of certain testing goal, such as fault detection. Current TCP research focus mainly on code-based regression testing; however in the Model-Based Testing (MBT) context, we still need more investigation. 
General TCP techniques do not use historical information, since this information is often unavailable. Therefore, techniques use different sources of information to guide prioritization. We propose a novel technique that guides its operation using provided hints, the Hint-Based Adaptive Random Prioritization - HARP.
Hints are indications or suggestions provided by developers about error-prone functionalities. As hints may be hard to collect automatically, we also propose an approach of collecting them. 
To validate our findings, we performed an experiment measuring the effect of introducing hints to HARP. It shows that hints can improve HARP's performance comparing to its baseline. Then, we investigated the ability of developers/managers to provide good hints and used them in a case study. This analysis showed that developers and managers are able to provide useful hints, which improves HARP's fault detection comparing to its baseline. Nonetheless, the provided hints should be consensual among the development team members.
\end{abstract}


\begin{multicols}{2}
\normalsize

\vspace{-4pt}
\section{Introduction}
\label{sec:introduction}
\vspace{-2pt}

Verification and Validation (V\&V) activities play an important role during software development. They often help to decrease the chances of delivering a buggy product and to increase the quality of software artifacts. Therefore, a great deal of the project's budget and resources is often devoted to V\&V tasks \cite{sommerville1}. One of these tasks is \textit{Software Testing}. When testing, a tester executes a program providing a set of controlled inputs, and checks the outputs, looking for errors, anomalies, and/or non-functional information~\cite{sommerville1}. Although widely used and extensively studied, software testing is known to be costly. Beizer \cite{beizer2}, and Ammann and Offutt \cite{AmmannO2008} point that testing alone can consume nearly 50\% of a project's budget. Therefore, research has been conducted aiming at reducing the costs related to software testing. 

In this scenario, the Model Based Testing approach (MBT)~\cite{utting1} has emerged as an alternative to traditional testing. MBT uses the system's models to automatically perform several testing tasks (e.g., test case generation). Among the main advantages of using MBT, we can list~\cite{utting1}: i) \textit{time and costs reduction} - automation is one of the pillars of MBT, which ends up helping the development team to save time and resources when building testing artifacts; and ii) \textit{artifacts soundness} - as test artifacts are derived automatically from system's models, it decreases the chances of creating incorrect artifacts. 

Due to the exhaustive nature of most test case generation algorithms used in MBT (e.g., based on depth-first search), generated test suites tend to be massive, which may jeopardize test case management~\cite{utting1}. To cope with this problem, a number of techniques have been presented in the literature. Most of them can be classified among three categories:  Test 
Case Selection~\cite{HemmatiAB13, emanuela1, harrold2}, Test Suite Reduction~\cite{CoutinhoCM14, SampathB12}, and Test Case Prioritization (TCP)~\cite{KaurBS12, jiang1, elbaum3}.

The goal of test case selection and test suite reduction is to propose a subset of test cases from the original suite, in other words, to help testers to work with a smaller amount of test cases. However, some test cases that unveil faults not detected by any other one may be discarded~\cite{jeffrey1}. Therefore, in some scenarios, both test case selection or reduction may fall short on proposing an effective smaller test suite that still reveals all faults. On the other hand, the TCP problem is to determine a permutation for the elements of a test suite that minimizes an evaluation function, chosen according to a prioritization objective \cite{elbaum3}. As a direct consequence, a test case prioritization technique does not eliminate any test case from the original test suite, but it helps to minimize resources usage by focusing on the most relevant ones~\cite{RUCH99}. By not discarding test cases, TCP is flexible, since it allows the tester to decide the amount of test cases to be run. This decision is often taken according to project's resources availability. Thus, TCP has been widely studied and is found as a good alternative to address the ``huge suite'' problem. 

Both code-based and model-based test suites can be prioritized using TCP techniques. However, most techniques presented in the literature have been defined and evaluated only for code-based suites, and in the context of regression testing \cite{elbaum3, jiang1}. Catal and Mishra \cite{CatalM12} conducted a systematic literature mapping regarding TCP and, among the results, they suggest that the amount of new model-based prioritization techniques has increased in a slower rate when compared to code-based ones. Moreover, they encourage novel works in this field.

To the best of our knowledge, there are only few attempts for defining TCP techniques based on model information. Korel et al.~\cite{KKT08} proposed a set of heuristics for TCP using Extended Finite State Machines - EFSM. On the other hand, Sapna and Mohanty~\cite{SapnaM09}, Kaur et al.~\cite{KaurBS12}, Stallbaum et al. \cite{StallbaumMP08}, and Nejad et al.~\cite{NejadAD16} proposed TCP techniques that prioritize test cases generated from UML Activity Diagrams. In addition, Kundu and Sarma~\cite{KunduSSM09} proposed a technique suitable for UML Sequence Diagrams. However, effectiveness and applicability of current techniques still require further investigation.

In a previous work~\cite{OuriquesJSERD}, we performed a set of empirical studies evaluating how several factors, such as number of faults, elements of model's layout (represented by the amount of branches, joins, and loops it has), and failing test cases' characteristics, can affect the behavior of traditional TCP techniques in the MBT context, focusing on Labeled Transition Systems (LTS). Our results showed that, for models with different layouts, the investigated techniques did not present variation in fault detection capability, which indicates low effects of model layout. Besides, the study pointed out that characteristics of failing test cases (such as the amount of branches, joins and loops they traverse in the model) can explain techniques' performance. Therefore, these studies suggested that TCP techniques should not rely only on layout characteristics. As secondary result of these studies, we also observed that techniques implementing the Adaptive Random strategy \cite{jiang1, Zhou10} presented a good performance, mainly because they have the adaptive component, exploring more evenly the test cases. However, they were also negatively affected by random choices, which was exposed by spread boxplots.

The effectiveness of a prioritized test suite is often evaluated by its ability of revealing faults as fast as possible. Ideally, a technique should place all test cases that unveil faults in the first positions of the prioritized test suite. However, this may require key information that may not be available, such as historical data about previous test executions, as in the Regression Testing context~\cite{CatalM12}. On the other hand, general TCP techniques do not take into account historical data. They can be very useful during the initial cycles of system testing since a great amount of never executed test cases are often available, and there are only preliminary information to guide the prioritization process \cite{LuLCZHZZ16}. There are some techniques that rely on the experts' knowledge as a guide to the prioritization task \cite{StallbaumMP08, YooHTS2009, TonellaAS2006}, but they demand a high interaction with the specialist, which increases their application costs. In this work, we focus on the general TCP problem in the system level context, considering test suites generated through MBT approaches, assuming that historical information from previous testing executions is not available, and using expert's knowledge to leverage the prioritization process.

Motivated by the lack of confidence of taking into account the model layout itself and the promising performance of techniques based on the adaptive random strategy, we propose \textbf{Hint-Based Adaptive Random Prioritization - HARP}. It comprises two main steps: the acquisition of \textbf{hints} from developers and managers, and the prioritization of test cases using these hints as a guidance layer. By ``hints" we mean indications from developers and managers -- which we refer to as \textit{development team} -- regarding jeopardized portions of the system, for instance, portions of the implemented use cases that were hard to implement or even subject to changes in schedule. HARP's objective is to take advantage of the expertise acquired during the system development to improve the exploration of test cases and consequently help to better order the test cases for anticipate fault detection.

HARP's first step aims at collecting hints from the development team with respect to the use cases that they actually worked on through a questionnaire. It is important to remark that hints about error-prone functionalities could be obtained from a number of different sources, for instance change impact analysis data, defect prediction, and risk assessment \cite{XuLYAJ16, RadjenovicHTV13}, provided that history on test execution is required. However, this investigation is out of the scope of this paper, since we are not addressing the regression testing context. 

Using the hints as a layer of guidance, HARP takes as input a test suite and suggests a new order for their elements. Each test case is a sequence of labels representing \textbf{steps} performed by the tester, \textbf{results} that the system under testing (SUT) should produce, or \textbf{conditions} that must be satisfied to proceed executing the test case. A hint is a piece of information from the development team, which is translated to a test purpose that, in turn, is an expression defined from test case labels, that filters the set of available test cases. Consequently, HARP is independent of specific kinds of models as well as of the test case generation process.

We evaluate \textbf{HARP} through three empirical studies: 
an experiment evaluating the affect of giving good and bad hints on HARP;
a questionnaire with development teams from two industrial projects, evaluating whether the participants would be able to provide proper hints, i.e., it evaluates whether the use case steps provided by the participants are actually related to faults;
and a case study comparing HARP with the original ARP technique \cite{jiang1}, involving test suites from industrial systems, using the hints collected in the questionnaire. 

Our experiment suggested that HARP gets to its best performance when one provides a good hint, which means that even with the random aspects, the algorithm benefits from hints; on the other hand, if the provided hint is bad, HARP is negatively affected and should not be used in these scenarios. Since we are suggesting that a particular hint to be used only whether the majority of the developers and managers agree about it, we believe that scenarios where a hint would misguide the prioritization are being minimized.
From the questionnaire we found out that, in most of the cases, participants were able to indicate error prone paths, which evidences the real applicability of our technique. 
From the case study we found out that, for the investigated applications and guided by the hints provided through the questionnaire, HARP reduced the time to reveal the first fault in comparison to traditional ARP.

In summary, this work has the following contributions: i) a novel TCP technique -- HARP -- along with its algorithm and asymptotic analysis; ii) a way to collect helpful information to guide prioritization from development teams; iii) an experiment exposing the effects of providing a hint to the provided algorithm; and iv) a case study comparing the ability of revealing the first fault of HARP and its baseline. In addition, we provide enough detail to stimulate a repetition of this investigation, as well as scripts and collected data, providing transparency and reproducibility to the investigation.

The remainder of the paper is structured as follows: 
Section~\ref{sec:background} exposes the background topics involved in this work;
Section~\ref{sec:example} depicts a motivational example introducing the idea of the hint and how to obtain it; Section~\ref{sec:technique} details the proposed technique and provides an asymptotic analysis; 
Section~\ref{sec:investigation} presents details about the conducted empirical investigation; 
Section~\ref{sec:relatedwork} presents the related work; 
Finally, Section~\ref{sec:conclusions} summarizes the conclusions about the work.
\vspace{-4pt}
\section{Background}
\label{sec:background}
\vspace{-2pt}

This section introduces the key concepts used in this work. It covers the formal definition of the Test Case Prioritization problem (Section~\ref{sec:TCP}), the Adaptive Random strategy (Section~\ref{sec:ART}), the activity of modeling a system (Section~\ref{sec:modeling}), and the use of Test Purposes on testing techniques (Section~\ref{sec:TP}).

\subsection{Test Case Prioritization}
\label{sec:TCP}	

Test Case Prioritization (TCP) techniques reorder the test cases from a test suite in order to quicken the reaching of a given testing objective. Testers may apply TCP on both code-based and model-based contexts, but it has been more frequently used in the code-based context and often related to regression testing~\cite{CatalM12}. Formally, the TCP problem is defined as following \cite{ElbaumMR00}:

\begin{mydef}[TCP Problem]
\textit{Given:} A test suite $TS$, the set of all permutations of $TS$, $PTS$, and a function from $PT$ to real numbers $f : PTS \rightarrow \mathbb{R}$.
\\
\textit{Problem:} To find a $TS'$ $\in$ $PTS$ $\mid$ $\forall$ $TS''$ ($TS''$  $\in$ $PTS$) ($TS''$ $\neq$ $TS'$) $\cdot$ $f$($TS'$) $\geq$ $f$($TS''$).
\end{mydef}

In other words, for the set of all permutations of the available test cases, the problem is to select one that maximizes the evaluation function, i.e. a function that measures how good is a prioritized test suite.

TCP has two clear limitations: i) to analyze the whole set of all permutations of $TS$; and ii) the evaluation function $f$. Since the amount of permutations of a set $A$ is $|A|!$, whether $TS$ is big, analyzing the elements from $PTS$ might be impractical, as discussed by Lima et al.~\cite{LimaISA2009}. Thereby, usually TCP techniques aim at selecting one test case at a time and place it in the desired order following certain heuristics. Moreover, the evaluation function is related to the prioritization goal, for example, to reduce test case setup time \cite{LimaISA2009} and, more commonly, to accelerate fault detection. Depending on the prioritization goal, the required information to define $f$ is not available beforehand, making its definition impossible. For instance, considering fault detection, the information regarding where the faults are or which test cases detect them is not available in advance, then a technique is not able to maximize it. Therefore, techniques that focus on fault detection, estimate fault information through surrogates \cite{YooHTS2009}.

\subsection{Adaptive Random Strategy}
\label{sec:ART}

Chen \textit{et al.} \cite{ChenLM04} propose the Adaptive Random strategy as an alternative to pure random test case generation for software with numerical inputs. The authors introduce the term \textbf{failure pattern}, which is a region in the input space that causes failures. Thus, the adaptive random strategy lies in estimating these failures patterns and spread the test cases more efficiently through the input space. The original strategy is suitable for on-line test case generation with numerical inputs \cite{ChenLM04}. After a test case is run, its result determines whether the algorithm should generate another one, i.e., if a test case does not reveal any fault, the strategy generates another test case, until a failure occurs. The idea is to randomly sample a set of candidate test cases and, among them execute the \textit{farthest} away from the already executed ones. This process continues until the execution unveils a fault.

The distance concept, represented by the \textit{farthest away} expression, is a set of functions that evaluates two test cases and two sets of test cases. The first one calculates the distance between two test cases, taking into account the estimated failure pattern. Chen \textit{et al.} \cite{ChenLM04} propose the use of the Euclidean Distance function to evaluate two test cases that manipulate numerical inputs. To compare two sets of test cases, the algorithm looks for the maximum of the minimum distances between the already executed test cases and the candidates.

Jiang \textit{et al.} \cite{jiang1} propose the use of the adaptive random strategy in order to solve the general TCP problem. For that, they propose a technique, which we refer to as the Adaptive Random Prioritization ({\bf ARP}) technique, by adapting the original strategy as follows: i) turning the \textbf{executed set} into a \textbf{prioritized sequence}, and ii) changing the stop criterion to \textbf{repeat the process until all test cases are in the prioritized sequence}. Thus, the proposed technique places test cases in order while the untreated set of test cases is not empty. The authors use Jaccard \cite{Jaccard01} as distance function, which measures the distance between sets by comparing the union and intersection sets. Later, Zhou \cite{Zhou10} suggests another use for the previously proposed technique. The author changes the distance function to the Manhattan function, but keeps the same algorithm, which compares two test cases through their branch coverage. Moreover, Zhou et al.~\cite{ZhouSS12} perform an empirical study comparing these two variations. The results of this study show that, in the code-based context, the Manhattan distance provides better fault detection. This fact is mainly because it measures just coverage of code elements, instead of measuring the frequency of the coverage, as it is done by the Jaccard function.

\subsection{Modeling Systems}
\label{sec:modeling}

In Model-Based Testing, the main artifact is the system model, which often represents its behavior through execution flows. The first decision is to define the model notation based on the abstraction level and the testing level. In this work, since we consider a higher abstraction level in system level, we consider Transition Based models due to their high abstraction, wide use, availability of supporting tool, and its suitability for control-flow representation \cite{utting1}. In our work, we are using essentially Labeled Transition System - LTS as system models. An LTS is defined in terms of states and labeled transitions linking them. Formally, an LTS is a 4-tuple $<S, L, T, s_0>$, where \cite{VriesT00}:

\begin{itemize}
	\item $S$ is a non-empty and finite set of \textit{states};
	\item $L$ is a finite set of \textit{labels};
	\item $T \subseteq SxLxS$ is a set of triples, the \textit{transition relation};
	\item $s_0$ is the \textit{initial state}. 
\end{itemize}

\begin{figure}
	\footnotesize\centering
	\centerline{\includegraphics[width=0.9\linewidth]{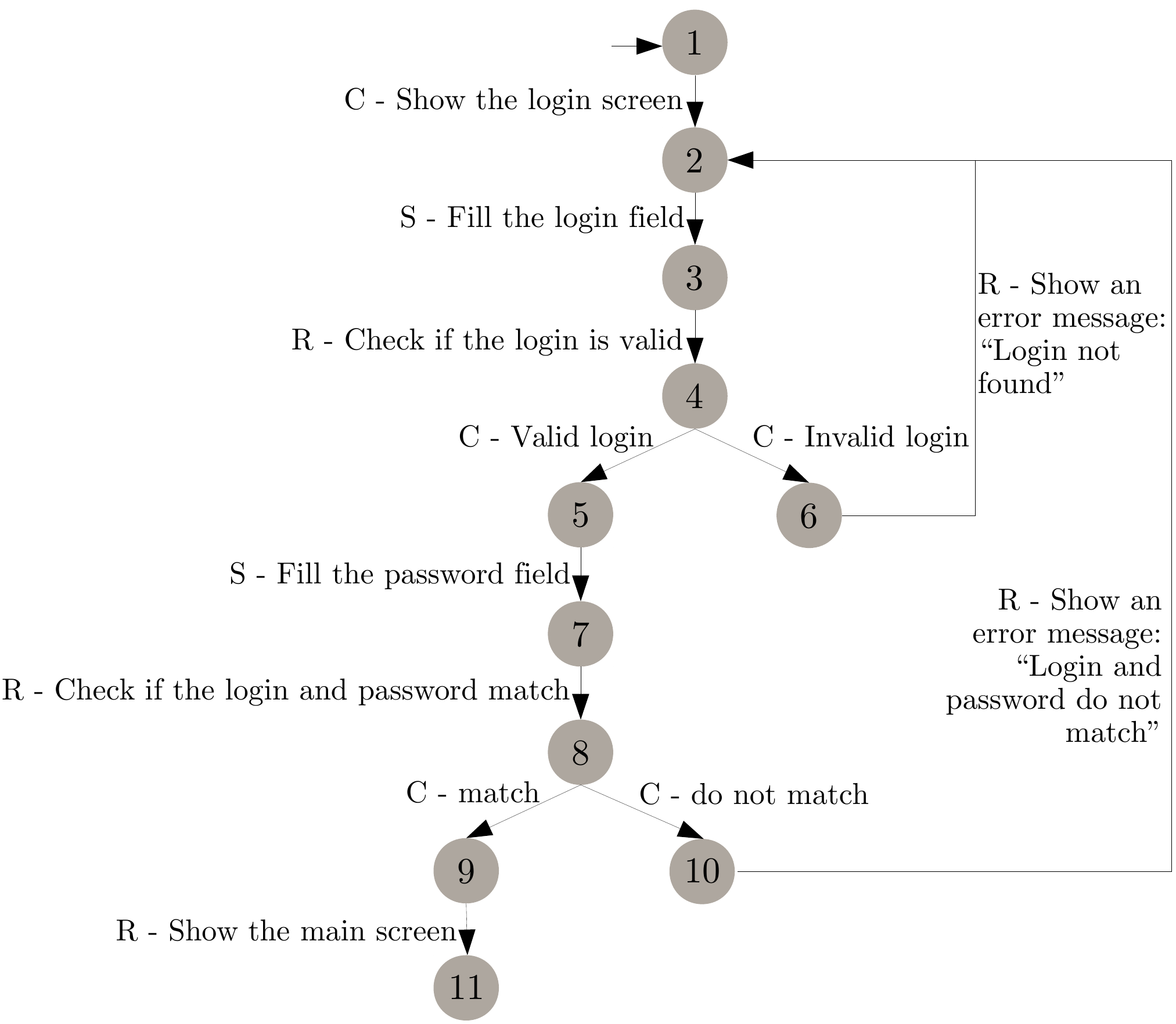}}
	\caption{Model representing an use case for login and password verification.}
	\label{fig:ExampleModel}
\end{figure}

To model a system using this notation, one needs to represent the \textbf{user steps}, \textbf{system responses}, and \textbf{conditions} as LTS transitions and to represent a decision, one can just branch the state and add a transition representing each alternative. For the sake of notation, the transition's label has a dedicated prefix: \textit{``S - "} for user steps, \textit{``R - "} for system responses, and \textit{``C - "} for conditions. It is also possible to represent the junction of more than one flow, and loops. One can represent three kinds of \textbf{execution flows} using this notation: \textit{base}, describing the main use of the system, i.e. the scenario where the user does the expected in most situations and no error occurs; \textit{alternative}, defining an user's alternative behavior or a different way some action can be done; and \textit{exception}, specifying the occurrence of an error returned by the system. In Figure~\ref{fig:ExampleModel} we can see an example of an application that verifies login and password, modeled through an LTS. The first transition represents the system precondition, which is when it shows its main screen. It can be interpreted as an initial condition for the system operation. In the next step, the user must fill the login field and then, the system verifies if the filled login is valid. Depending on the login validity, the execution follows two different paths, expressed by the branches going out from the state labeled as ``4''. If the login is invalid, a loop transition that shows an error message takes the execution flow back to the action of filling the login field.

\subsection{Test Purpose}
\label{sec:TP}

Test purpose is the purpose of testing! Indeed, a test purpose indicates the intention behind a testing process. According to Vain \textit{et al.} \cite{VainRKE07}, a test purpose is a specific objective or property of the system under test that the development team wants to test. In practice, applying a test purpose to a test suite represents a filtering process, resulting in a subset of test cases that satisfies the purpose.

There are different ways for representing test purposes. Jard and Jeron~\cite{Jeron2005} present a formal approach for representing test purposes in conformance testing. This representation uses an extended version of Labeled Transition Systems (LTS), which includes input and output actions (IOLTS). The same format is suggested to be used for modeling the system under testing. Therefore, with the system and the test purpose respecting the same rules, the latter is able to guide the testing process through the former.

Cartaxo et al.~\cite{ltsbt08} proposes a test purpose also based on LTS models but considering the ``accept'' or ``reject'' modifier. This notation is more flexible, since it does not need to define specific inputs and outputs, just the steps of the model represented by the edges of the LTS. Moreover, this representation can be also applied in the MBT context by considering system level test cases. This test purpose is expressed using the following guidelines:

\begin{itemize}
  \item The purpose is a sequence of strings separated by commas;
  \item Each string is a label of an edge from the model or an ``*'', which works as a wildcard, indicating that any label satisfies the purpose;
  \item The last string is ``accept'', if the test case must agree with the test purpose definition, or ``reject'' otherwise.
\end{itemize}

HARP takes as input a test purpose, which format is a variation of Cartaxo's test purpose~\cite{ltsbt08}. We adapted it by omitting the last string, which is \textit{accept} or \textit{reject}, adopting just the acceptance. This modification was needed because we focus only on hints regarding error-prone regions of the system, regardless of safer portions. Therefore, our test purpose is defined as follows:

\begin{itemize}
	\item A sequence of strings separated by a vertical bar or pipe (``$|$''); 
	\item Each string is the label of an element of the system model, e.g. a step of a use case, or a ``*'', which is a wildcard, representing any text.
\end{itemize}
\vspace{-4pt}
\section{Motivating Example}
\label{sec:example}
\vspace{-2pt}

We measure the success of a TCP technique that focus on fault detection by analyzing whether it creates prioritized test suite orders that anticipate fault detection, when compared to the original suite. Therefore, a TCP technique ideally places all test cases that fail in the first positions of the prioritized test suite. However, this may require key historical information that is often not available. Considering the general TCP context, some techniques use as prioritization criteria model characteristics such as structural aspects of the models. For instance, a greedy prioritization technique \cite{elbaum3} would reschedule test cases according to the amount of steps each one describes. Therefore, it assumes that the more steps a test case has, more likely it is to reveal a fault.

To illustrate the problem addressed in our work, consider the LTS model depicted in Figure~\ref{fig:ExampleModel}, which specifies a use case that performs a login and password verification in an information system. Now, suppose that a tester applies a test case generation algorithm that traverses loops at most twice and this process creates a test suite with seven test cases (Table~\ref{tab:exampletestcases}). 

Now, suppose that a fault occurs when the user provides an invalid login twice in a row. Considering this scenario, \textbf{TC7} is a test case that covers the respective path of the model, and consequently should reveal the system's fault.

\begin{table}[h]
\centering
\footnotesize
\caption{Generated test cases from the example model. We use the notation based on the node's labels for the sake of visualization.}
\label{tab:exampletestcases}
\begin{tabular}{|c|c|}
\hline Label & Test Case Nodes\\ 
\hline TC1 & [1, 2, 3, 4, 5, 7, 8, 9, 11] \\ 
\hline TC2 & [1, 2, 3, 4, 5, 7, 8, 10, 2, 3, 4, 5, 7, 8, 9, 11] \\ 
\hline TC3 & [1, 2, 3, 4, 5, 7, 8, 10, 2, 3, 4, 5, 7, 8, 10, 2] \\ 
\hline TC4 & [1, 2, 3, 4, 5, 7, 8, 10, 2, 3, 4, 6, 2] \\ 
\hline TC5 & [1, 2, 3, 4, 6, 2, 3, 4, 5, 7, 8, 9, 11] \\ 
\hline TC6 & [1, 2, 3, 4, 6, 2, 3, 4, 5, 7, 8, 10, 2] \\ 
\hline TC7 & [1, 2, 3, 4, 6, 2, 3, 4, 6, 2] \\ 
\hline 
\end{tabular} 
\end{table}

With that in mind, suppose that the tester intends to accelerate fault detection by applying test case prioritization. Therefore, he/she decides to apply the aforementioned greedy strategy for rescheduling his test suite. After running it, he/she collects the following prioritized test suite: $PTS_G = [TC2, TC3, TC4, TC5, TC6, TC7, TC1]$. As the greedy strategy considers only \textit{number of steps} as prioritization criteria, it ended up placing the single test case that reveals the fault at the penultimate position, therefore representing an undesirable order for a TCP technique.

Now, suppose that during the development of the referred use case, due to time restrictions, the team end up, by time restriction, neglecting the verification of the scenarios where the login and password are not valid. Or else, also during development, the team can perceive that the login and password verification is not a trivial feature, e.g., login and passwords must adhere to special patterns, or rely on external resources, for example an unstable network link. In both cases, take into account this information could potentially lead to an indication of use case error prone regions. Therefore, a developer may provide a hint, for example, ``Invalid login attempts may not have been properly handled due to resource constraints", that can be translated by a tester into a test purpose as \textbf{``* $|$ C - Invalid Login $|$ *"}, indicating that any test case that cover a scenario where the provided login is invalid, would be interesting to test. Thus, a hypothetical prioritization technique that considers this hint would place test cases related to these problematic paths in the suite's first positions. For instance, test cases that cover the steps where the login is verified and an error occurs would be run first. Among the possible output test suites for this hint strategy can be $PTS_H = [TC4, TC7, TC5, TC6, TC3, TC2, TC1]$. As we can see, \textbf{TC7} is placed at the beginning of the prioritized test suite, which is a more desirable performance for a TCP technique.

We believe it is valid to assume that the developers' expertise acquired during the development of certain functionality can be very useful for guiding the prioritization process, since he/she often owns key information regarding problematic and/or complex software features/modules, which can lead to fault inclusion. Therefore, we propose a technique that takes into account hints provided by the development team and uses them as estimators of error prone portions of the system.
\vspace{-4pt}

\section{Hint-Based Adaptive Random Prioritization}
\label{sec:technique}

\vspace{-2pt}

We propose HARP as an adaptive random prioritization technique that uses hints from developers and managers to prioritize test cases from a test suite. In this section we detail both steps that it comprises.

\subsection{Hint Acquisition}

Hints are indications that the development team give - using use case documents about occurrences regarding the system under development/testing - that may contribute to the insertion of faults, such as portions of the system that some developer considered hard to implement, that uses some external and/or untrusted library, or that suffered a schedule shortening. Our hypothesis is that the aforementioned occurrences are estimators for faults and taking them into consideration should lead to better fault detection capabilities.

We designed HARP to process hints encoded as test purposes using the notation presented in Section~\ref{sec:TP}. Encoding hints as test purposes is a manual, but simple task. A single hint may be a single step in a use case or a sequence of steps (or even a whole flow). In the first case, suppose that the development team indicate a step with text ``single step" as hint; so the equivalent test purpose is $tp_1 = $``* $|$ single step $|$ *", indicating that any test case that traverse ``single step" is filtered by $tp_1$. On the other hand, in the second case, let the hint comprises a pair of step and expected result with texts ``single step" and ``single system response" respectively; the equivalent test purpose is $tp_2 = $``* $|$ single step $|$ single system response $|$ *", indicating that any test case that describes the referred step and provides expects the related response is filtered by $tp_2$. The same idea applies for a whole flow. Therefore, the output of this step is a set of hints $TP=\left\lbrace tp_1, tp_2, \cdots, tp_n\right\rbrace $ translated into test purposes. 

As illustration, considering the example in Section~\ref{sec:example}, suppose that the testers suspect that the error messages are not correct or are being acquired using an error-prone strategy. Therefore, they could codify hints as the set 
$TP_{ex}=\lbrace ``* | \textrm{ C - Invalid Login } | *", ``* | \textrm{ C - do not match } | *"\rbrace$.
The set contains two hints, each one about a scenario where the SUT prints an error message.

\subsection{Test Case Prioritization}

HARP is targeted for MBT test suites, potentially for manual execution, where test cases are expressed as sequences of steps, but do not rely on any specific type of model or test case generation algorithm. It combines the advantages of exploring evenly the test suite with the guidance provided by hints in order to improve the performance of an adaptive random-based technique, since it reduces random choices.

By observing the original adaptive random prioritization algorithm~\cite{jiang1, Zhou10}, we identified three points that could be modified: i) the selection of the first test case to be placed in order; ii) the generation of a candidate set of test cases to be placed in order in a particular algorithm's iteration; and iii) the functions that represents the notion of resemblance\footnote{We use the term \textit{resemblance} because in the literature there are two different notions, which are \textbf{distance} or \textbf{similarity}. Authors suggest functions following these two notions but both may be applied here.} and that selects the next test case among the candidates to be placed in order. In this work, we modify these three aspects in order to reduce the effects of random choices and to take into consideration the hint-based guidance. Algorithm~\ref{alg:HARP} summarizes the HARP technique. The method \texttt{prioritize}, which is the main function, takes as input a test suite $U\_TS$ and a set of test purposes $TP$, and returns a prioritized test suite.

\begin{algorithm}[H]
\caption{The main procedure.}
\label{alg:HARP}
\begin{algorithmic}[1]
	\Function{Prioritize}{$U\_TS$, $TP$}
		\State $PTS \gets \emptyset$
		\State $filtTCs \gets \textbf{filter(U\_TS, TP)}$
		\State $first \gets \textbf{firstChoice(filtTCs)}$
		\State $PTS.append(first)$
		\State $U\_TS.remove(first)$
		\State $filteredTCs.remove(first)$
		\While{$U\_TS \neq \emptyset$}
			\State $c \gets \textbf{genCandSet}(U\_TS, filtTCs)$
			\State $nextTC \gets \textbf{selectNext(PTS, c)}$
			\State $PTS.append(nextTC)$
			\State $U\_TS.remove(nextTC)$
			\State $filtTCs.remove(nextTC)$
		\EndWhile\\
		\Return $PTS$
	\EndFunction

\end{algorithmic}
\end{algorithm}

The algorithm starts with an empty set of test cases (Line 2). In Line 3, it filters $U\_TS$ using every test purpose from $TP$ and the remainder of the algorithm uses the result set as hint-related test cases. In Line 3, the algorithm defines the first test case to be placed in order (\texttt{contextAwareChoice}). For that, it selects randomly one among the filtered set. This selected test case is then included in the prioritized sequence (Line 4), and removed from the original set (Line 5) and from the filtered set (Line 6). 

The loop in lines 8-14 assembles the rest of the prioritized suite. To do so, in its first step, it generates a set of candidates (Line 9). The candidate set generation process selects randomly one test case at a time, while the set keeps increasing branch coverage and size less than or equals to 10, as discussed by Jiang et al. \cite{jiang1} and empirically evaluated by Chen \textit{et al.} \cite{ChenLM04}. We modified the candidate set generation by increasing the chances of hint-related test cases not yet placed in order be selected. In order to decide whether the next test case to compose the candidate set comes from the hint-related ones or not, we define a random variable following uniform distribution $V \sim U(0, 1)$. Therefore, iteratively, if the algorithm samples a value $v < 0.5$, it selects randomly a test case among the ones filtered by the test purposes, otherwise selects randomly a test case among the ones not yet prioritized but not related to the hints. 

The next step is to select the next test case to be placed in order among the candidates by calling \texttt{selectNextTestCase} (Line 10). Algorithm \ref{alg:HARPP2} gives details on how it works. First, it builds a matrix (Line 5) that represents the resemblance between candidates and the test cases that were already prioritized. Since we use the notion of similarity to represent the resemblance between test cases, the matrix contains similarity values. Then, the algorithm selects the candidate with the highest similarity (Line 8), i.e. the next test case will be the most similar to the previous prioritized ones. In order to measure the similarity between test cases we use the similarity function proposed by Coutinho et al. \cite{CoutinhoCM14}, which was proposed specifically to the MBT context, being fully compatible with test cases generated with different algorithms, since it takes into account, besides the common transitions, their frequency in the test cases. The function is defined by: $similarity(i, j) = \dfrac{nip(i, j) + |sit(i, j)|}{(\dfrac{|i| + |j| + |sdt(i)| + |sdt(j)|}{2})}$, where:

\begin{itemize}
	\item $i$ and $j$ are two test cases;
	\item $nip(i, j)$ is the number of identical transition pairs between two test cases;
	\item $sdt(i)$ is the set of distinct transitions in the test case $i$;
	\item $sit(i, j)$ is the set of identical transitions between two test cases, i.e. $sdt(i) \cap sdt(j)$.
\end{itemize}

\begin{algorithm}[H]
\caption{The select next test case procedure}
\label{alg:HARPP2}
\begin{algorithmic}[1]
	\Function{selectNextTestCase}{$p\_TS, c\_S$}
		\Comment{p\_TS is the already prioritized test sequence}
		\Comment{c\_S is the candidate set}
		\State $d \gets array[p\_TS.size][c\_S.size]$
		\For{i = 0 to p\_TS.size - 1}
			\For{j = 0 to c\_S.size - 1}
				\State $d \gets \textbf{sim(p\_TS[i], c\_S[j])}$
			\EndFor
		\EndFor
		\State $index \gets \textbf{maxValue(d)}$
		\State $nextTestCase \gets c\_S.get(index)$\\
		\Return $nextTestCase$
	\EndFunction
\end{algorithmic}
\end{algorithm}

\subsection{Running Example}

In this section, we provide a visual and practical understanding on how HARP works with an example. To do so, consider the scenario introduced in Section~\ref{sec:example}, in which we present a behavioral model of a fictitious application for login and password verification, and Table \ref{tab:exampletestcases} shows its generated test suite. 

Besides, suppose that the development team responsible for building this system identifies the verification of the login in the database as an error-prone step of the system, particularly the invalid login verification. This step may be risky because it verifies a non-encrypted data from the database and, if the developer did not think carefully when coding it, he/she might have left an open window for security attacks (e.g., SQL injection). Therefore, the tester formulates the hint as a set with a single test purpose $logError = $\{``* $|$ [C - Invalid login] $|$ *"\}. 

Considering the original test suite (refer to Table~\ref{tab:exampletestcases}), to propose a new execution order, HARP chooses the test case to be placed in the first position by filtering the input test suite according to the provided hint. The algorithm filters the test suite by visiting every test case, traversing it and verifying whether its sequence of the steps matches the test purpose, a process similar to the evaluation of a regular expression. The filtered subset is $filtered=\{TC4, TC5, TC6, TC7\}$. Then, suppose that the first test case randomly selected from the filtered set is \textbf{TC6}. Therefore, the algorithm removes it from the untreated test suite and makes $prioritizedSequence = [TC6]$.

Then, HARP creates a candidate set by randomly and iteratively selecting test cases from the untreated test suite or from the hint-related set. Suppose that the set is $candidateSet = \{TC1, TC3, TC5\}$, where just $TC5$ comes from the hint-related set. The next step is to calculate the similarities between the elements from current prioritized sequence and the ones from the candidate set. The calculated values are in Table~\ref{tab:similarities1}.


\begin{table}[h]
\centering
\footnotesize
\caption{1st Iteration: Similarities among candidates and test cases already prioritized}
\label{tab:similarities1}
\begin{tabular}{|l|l|}
\hline
             & \textbf{TC6} \\ \hline
\textbf{TC1} & 63.15\%      \\ \hline
\textbf{TC3} & 71.11\%      \\ \hline
\textbf{TC5} & 72.72\%      \\ \hline
\end{tabular}
\end{table}

Considering these values, \textbf{HARP} selects the test case with the highest similarity, \textbf{TC5}, then the algorithm adds it to the prioritized sequence, making $prioritizedSequence = [TC6, TC5]$. 

Now, in the second iteration of the main loop of the algorithm, it proceeds by generating a new candidate set, and suppose the second candidate set is $candidateSet = \{TC2, TC4\}$, where $TC4$ comes from the hint-related set. In the next step, it calculates the similarities between the test cases from $candidateSet$ and $prioritizedSequence$, as can be seen in Table~\ref{tab:similarities2}.

\begin{table}[h]
\centering
\footnotesize
\caption{2nd Iteration: Similarities among candidates and test cases already prioritized}
\label{tab:similarities2}
\begin{tabular}{|l|l|l|}
\hline
             & \textbf{TC6} & \textbf{TC5} \\ \hline
\textbf{TC2} & 59.57\%      & 68.08\%      \\ \hline
\textbf{TC4} & 90.90\%      & 72.72\%      \\ \hline
\end{tabular}
\end{table}

By comparing the maximum similarities for the two candidates, which are 68.08\% and 90.90\%, the next test case to be placed in order is \textbf{TC4}, making $prioritizedSequence = [TC6, TC5, TC4]$. The algorithm repeats this process until the last test case is placed in order. After the whole execution, adding one test case at a time, a proposed sequence could be $prioritizedSequence = [TC6, TC5, TC4, TC2, TC3, TC7, TC1]$.

\subsection{Asymptotic Analysis}

To investigate how costly it would be to run \textbf{HARP} in practice, we analyze its asymptotic bounds. In order to perform this analysis, we consider its worst execution scenario, which is when the maximum number of candidates (10), are selected at every iteration. Moreover, consider a test suite with $n$ elements and, for simplification, lets assume test cases from the original test suite have same size, $t$. The execution time for every step is as follows:

\begin{enumerate}
  \item \texttt{similarity} : The similarity calculation depends on the involved test cases sizes, however, as we are simplifying sizes to $t$, we can say its time is $2 \cdot O(t)$ or $O(t)$;
  
  \item \texttt{maxValue}: The definition of the maximum between the maximum similarities traverses a matrix with $<$number of candidates$>$ columns and $<$number of test cases already prioritized$>$ lines. The number of candidates will be in the interval of 1 and 10, and in the worst case, it will be 10 all iterations. The number of test cases already prioritized depends on the initial suite. Thus, the time is $O(10 \cdot n)$ or $O(n)$;
  
  \item \texttt{selectNextTestCase}: The selection of the next test case to be placed in order, once more, depends on the number of candidates and on the number of test cases already prioritized. The similarity is calculated for every pair of test cases and, following the same reasoning, the time is $10 \cdot n \cdot O(t)$ plus $O(n)$ of the \texttt{maxValue} execution. Thus, the execution time is $O(t \cdot n) + O(n)$ or $O(t \cdot n)$;
  
  \item \texttt{generateCandidateSet}: The generation of a candidate set depends on its size, that must be at most ten, and the increasing coverage of requirements. Considering the worst case, the time is constant, thus \textbf{$O(1)$};
  
  \item \texttt{contextAwareChoice}: The choice of the first test case depends on the amount of test cases from initial test suite and the time necessary to evaluate if a single test case matches the test purpose. The function just iterates over the initial test suite and verifies if each test case is accepted by the test purpose. This verification depends on the size of the test case and, since we assume that the size of the test cases are the same (equals to $n$), the verification is $O(t)$. Thus, the whole method executes in $n \cdot O(t)$ or $O(t \cdot n)$ time;
  
  \item \texttt{prioritize}: This function calls the above mentioned ones and its execution time is the total execution time of HARP. Thus, HARP execution time is $O(t \cdot n) + n \cdot (O(1) + O(t \cdot n)) = O(t \cdot n) + O(n) + O(t \cdot n^2)$, which is equals to $O(n^3)$, whether $n \geq t$ or $O(t \cdot n^2)$, otherwise.
\end{enumerate}
\vspace{-4pt}
\section{Empirical Investigation}
\label{sec:investigation}
\vspace{-2pt}

\textbf{HARP} relies on the assumption that development teams' members are able to provide good hints about functionalities they developed or helped to develop. Besides providing evidence of this assumption, we also intend to discuss the boundaries of HARP's applicability. In order to provide evidence about HARP's applicability and ability of revealing faults, we intend to investigate the following research questions:

\begin{itemize}
	
	\item \textbf{RQ1: Does the quality of a hint really affects HARP?} In this question we want to evaluate if a good hint leads to a significantly better ability of reveal faults in comparison with when receiving a bad hint, evaluating the effect size of HARP's performance equipped these two kinds of hints. We evaluate this to observe whether the random choices in HARP's algorithm affect the guidance provided by the hints;
	
	\item \textbf{RQ2: Are developers and managers able to provide good hints?} We are not interested in evaluating the participants itself. Instead, we try to find out whether the hints provided by them approximate portions of the system that	really contain faults;
	
	\item \textbf{RQ3: Is the hint collection process costly?} This information is important since we need to justify costs inserted by our technique, and it will show whether it can be applicable in practice. We are trying to find out how easy it can be to implement a process aiming to collect required information to derive the test purpose, i.e, the hints;
	
	\item \textbf{RQ4: IS HARP able to outperform the original Adaptive Random Prioritization technique, considering actual hints?} To address this question, we compare HARP with the baseline version of the Adaptive Random Prioritization proposed by Jiang et al. \cite{jiang1}.
\end{itemize}

To address these research questions, we perform three distinct empirical studies: an experiment controlling the quality of hints provided as input for HARP to address RQ1; a questionnaire involving developers and managers from a partner project addressing RQ2 and RQ3; and a case study involving industrial systems with hints provided by developers and managers to address RQ4. We report these studies in the next sections. This research has a companion site (https://goo.gl/FH3b5m) containing the collected data, R scripts for data analysis, as well as instructions on how to use them.

\vspace{-4pt}
\subsection{Experimental Study}
\label{sec:experiment}
\vspace{-2pt}

In this section, we report an experiment designed to investigate how HARP performs when good and bad hints are provided. Good hints are the ones that actually points to test cases that fail, on the other hand bad hints are the ones that do not point to test cases that fail. In order to guide this investigation and address RQ1 properly, we define a clear manual procedure to derive good and bad hints using actual fault reports and exercise HARP with them. Besides, as a further analysis, we also vary the distance function aiming at assessing the effect on them. Therefore, we define this experiment using the framework based on key concepts suggested by Wohlin \textit{et al.} \cite{wohlin1}:
		
\begin{itemize}
	\item The \textbf{objects of study} are \textbf{HARP}, \textbf{hints} and the \textbf{test suites} generated from models that represent the System Under Testing (SUT);
	\item The \textbf{purpose} of the study is to evaluate whether HARP is really affected by the hint's quality;
	\item The \textbf{quality focus} is the capacity of suggesting a test suite capable of revealing faults earlier;
	\item The \textbf{perspective} is from the tester point of view, i.e., the person that executes a TCP technique in a testing team;
	\item And the \textbf{context} is Model-Based Testing.
\end{itemize}
	
Based on the previous definition, we detail the experiment planning, which encompasses: the experiment context, its variables and experimental objects, experimental design, hypotheses under testing, and validity evaluation.

This is an \textit{offline} experiment, which comprises industrial applications, fault reports, and test cases generated through a test case generation algorithm. We define manually the artificial hints used in this study, since we need to represent both extremes of the situation: when the team provides either an accurate hint (one that points to test cases that fail) which we call a \textbf{good hint}, or a poor hint (one that does not point to any failure), which we call a \textbf{bad hint}. We define them with previous knowledge about faults and failures collected from fault reports provided during the development process.

\paragraph{Hint Definition}

To define a bad hint is straightforward; we define a test purpose with steps not related to any fault and make sure that any of the filtered test cases reveal a fault. On the other hand, a good hint must represent a scenario where the filtered test cases reveal a fault, whereas not biasing the investigation. The definition of a good hint for this study follows a systematic and manual approach:

\begin{itemize}
	\item For a specific fault of a given model, composed by its cause and the test cases that fail because of it, we define a test purpose with the label of the edge more related to the cause of the fault;
	
	\item If the whole set of test cases filtered by this test purpose fails, using it can bias the investigation, since a test case that fail would be chosen as the first one in sequence proposed by HARP. Therefore, we add more edges yet related to the cause in order to avoid this scenario;
	
	\item If it is not possible to create a single test purpose able to filter test cases that fail and that not fail, we define other test purpose related to the same hint and we restart the process aiming at filtering a set of test cases that does not bias the study.
\end{itemize}

As an example of this approach, consider the scenario suggested in our motivating example (refer Section~\ref{sec:example}). Assuming the same suggested fault, the hint definition process begins by creating a test purpose with the label of a single transition related to the fault, which is the one between states labeled as 4 and 6. The first row of Table~\ref{tab:hintGeneration} shows the filtered test cases and the proportion of the ones that fail considering a test purpose comprising the label of the transition more related to the considered fault. Since we know that the fault occurs when the user provide a wrong password twice in a row, we can be even more precise by adding the same label twice. The second row in the same table shows a test purpose that specifies the case. However, note that this test purpose filters only the test case that reveal the fault, which bias the study, because HARP would always place TC7 the in the first position in the prioritized test suite. Thus, according to the aforementioned approach, the test purpose `` * $|$ C - Invalid Login $|$ * " is the final one.

\begin{table}
	\centering
	\scriptsize
	\caption{Illustration of the good hint generation}
	\label{tab:hintGeneration}
	\begin{tabular}{|c|c|c|}
		\hline
		\textbf{Test Purpose}                                                                       & \textbf{Filtered TC} & \textbf{\% Failure} \\ \hline
		``*$|$C-Invalid Login$|$*"                                                                  & TC4, TC5, TC6, TC7   & 25\%                \\ \hline
		\begin{tabular}[c]{@{}c@{}}``*$|$C-Invalid Login$|$*\\ $|$C-Invalid Login$|$*"\end{tabular} & TC7                  & 100\%               \\ \hline
	\end{tabular}
\end{table}

We use the proportion of test cases that fail among the filtered ones to judge how good a hint is. Intuitively, for a bad hint 0\% of the filtered set fail, on the other hand, following the aforementioned approach to derive good hints lead to a proportion between 20\% and 50\%. These values represent the best test purposes we are able to generate using the aforementioned hint definition, considering the selected use cases and test reports, and without being extreme, i.e. neither filtering 100\% of the test cases and being equal to the baseline ARP technique, nor just one test case that unveil the fault. We use this approach to define the possible values for a variable investigated in this study.

\paragraph{Independent Variables}

Since we want to evaluate the effects of the hint's quality on HARP, observing variations by adopting different resemblance functions, we define the following independent variables:

\begin{itemize}
	\item Hint quality 
	\begin{itemize}
		\item Good hints (between 20\% and 50\% of the test cases filtered by hints fail)
		\item Bad hints (no test case filtered by hints fail)
	\end{itemize}
	\item Resemblance function 
	\begin{itemize}
		\item Similarity function \cite{CoutinhoCM14}(as proposed in Section~\ref{sec:technique})
		\item Jaccard distance function (as originally proposed by Jiang et al. \cite{jiang1})
	\end{itemize}
\end{itemize} 

\paragraph{Dependent Variable}

Our goal is to investigate whether hints affect HARP's results, regarding early fault detection. To help this investigation, we selected the APFD (Average percentage of Fault Detection) metric. The APFD is calculated as follows:

\begin{equation}
APFD=1-\dfrac{TF_1+TF_2+\ldots+TF_m}{nm}+\dfrac{1}{2n}
\end{equation}

\noindent where $TF_i$ is the position of the first test case that reveals the $i$-th fault, $m$ is the number of faults that the test suite is able to reveal, and $n$ is the size of the test suite \cite{elbaum3}. The APFD is a percentage in which a higher value indicates that the faults were revealed earlier during test execution and analogously, a lower value indicates that the faults were unveiled later.

\paragraph{Experiment Objects}

In the MBT process, the system's models are often the input artifacts. Test cases are generated from these models and later provided to prioritization techniques. As experiment objects, we use two industrial systems: the \textbf{Biometric Collector} and the \textbf{Document Control}\footnote{Developed as part of a cooperation between Ingenico do Brasil and our research lab}. Both systems were modeled through a set of Labeled Transition Systems, which represent their functionalities.

The test cases were generated by the testing team using the same test case generation algorithm, which traverses the LTS covering loops at most twice, and uploaded them to a test case execution management tool - Testlink\footnote{http://testlink.org/}. Then, we collected the test suites and used them in our experiment. To the purpose of our investigation, we consider only the ones that have at least one reported fault. The faults considered in the study were reported by testers also using Testlink. Thus, our models and faults are:

\begin{itemize}
	\item Biometric Collector
	\begin{itemize}
		\item CB01: which contains two faults, one revealed by two test cases, and the other by a single test case;
		\item CB03: which contains one fault revealed by a single test case;
		\item CB04: which contains four faults, two revealed by two test cases each, and the remaining two revealed by a single test case each;
		\item CB05: which contains three faults, one revealed by two test cases, and the remaining two revealed by a single test case each;
		\item CB06: which contains one fault revealed by a single test case. 
	\end{itemize}
	
	\item Document Control
	\begin{itemize}
		\item CD01: which contains two faults, one revealed by three test cases, and the other by a single test case;
		\item CD02: which contains one fault revealed by a single test case;
		\item CD03: which contains one fault revealed by a singe test case;
		\item CD04: which contains two faults, revealed by a single test case each;
		\item CD05: which contains two faults, one revealed by three test cases, and the other one by just a single test case.
	\end{itemize}
\end{itemize}

Since our objective was to provide a hint and detect a related fault earlier in the testing process, each experiment object was a pair $<testsuite,fault>$, where $testsuite$ is the test suite related to the selected use cases and $fault$ is a fault recorded by executing this test suite. For instance, for CB01, we considered as objects both $<$CB01,Fault1$>$ and $<$CB01,Fault2$>$. Thus, according to the list of models and faults exposed previously, we worked with 19 experiment objects. Although we are not able to provide the artifacts itself, Table~\ref{tab:modelInfo} exposes some metrics about them.	

\begin{table*}
	\centering
	\footnotesize
	\caption{Metrics about models used as experiment objects}
	\label{tab:modelInfo}
	\begin{tabular}{|c||ccccc|ccccc|}
		\hline
		\multirow{2}{*}{}  & \multicolumn{5}{c|}{\textbf{Biometric Collector}}                                                                     & \multicolumn{5}{c|}{\textbf{Document Control}}                                                                        \\
		& \textbf{CB01}           & \textbf{CB03}           & \textbf{CB04}           & \textbf{CB05}           & \textbf{CB06} & \textbf{CD01}           & \textbf{CD02}           & \textbf{CD03}           & \textbf{CD04}           & \textbf{CD05} \\ \hline
		\textbf{Branches}  & \multicolumn{1}{c|}{9}  & \multicolumn{1}{c|}{9}  & \multicolumn{1}{c|}{11} & \multicolumn{1}{c|}{17} & 4             & \multicolumn{1}{c|}{7}  & \multicolumn{1}{c|}{7}  & \multicolumn{1}{c|}{5}  & \multicolumn{1}{c|}{2}  & 9             \\ \hline
		\textbf{Joins}     & \multicolumn{1}{c|}{3}  & \multicolumn{1}{c|}{3}  & \multicolumn{1}{c|}{4}  & \multicolumn{1}{c|}{11} & 0             & \multicolumn{1}{c|}{4}  & \multicolumn{1}{c|}{4}  & \multicolumn{1}{c|}{5}  & \multicolumn{1}{c|}{2}  & 9             \\ \hline
		\textbf{Loops}     & \multicolumn{1}{c|}{2}  & \multicolumn{1}{c|}{2}  & \multicolumn{1}{c|}{1}  & \multicolumn{1}{c|}{6}  & 0             & \multicolumn{1}{c|}{2}  & \multicolumn{1}{c|}{2}  & \multicolumn{1}{c|}{0}  & \multicolumn{1}{c|}{1}  & 0             \\ \hline
		\textbf{Max Depth} & \multicolumn{1}{c|}{12} & \multicolumn{1}{c|}{20} & \multicolumn{1}{c|}{36} & \multicolumn{1}{c|}{32} & 14            & \multicolumn{1}{c|}{22} & \multicolumn{1}{c|}{22} & \multicolumn{1}{c|}{12} & \multicolumn{1}{c|}{14} & 12            \\ \hline
		\textbf{Min Depth} & \multicolumn{1}{c|}{8}  & \multicolumn{1}{c|}{4}  & \multicolumn{1}{c|}{14} & \multicolumn{1}{c|}{12} & 6             & \multicolumn{1}{c|}{10} & \multicolumn{1}{c|}{10} & \multicolumn{1}{c|}{10} & \multicolumn{1}{c|}{14} & 10            \\ \hline
		\textbf{Gen. TC}   & \multicolumn{1}{c|}{16} & \multicolumn{1}{c|}{14} & \multicolumn{1}{c|}{16} & \multicolumn{1}{c|}{67} & 3             & \multicolumn{1}{c|}{10} & \multicolumn{1}{c|}{10} & \multicolumn{1}{c|}{5}  & \multicolumn{1}{c|}{3}  & 9             \\ \hline
	\end{tabular}
\end{table*}

\paragraph{Experiment Design}

\begin{figure}[h]
\footnotesize\centering
\centerline{\includegraphics[width=0.9\linewidth]{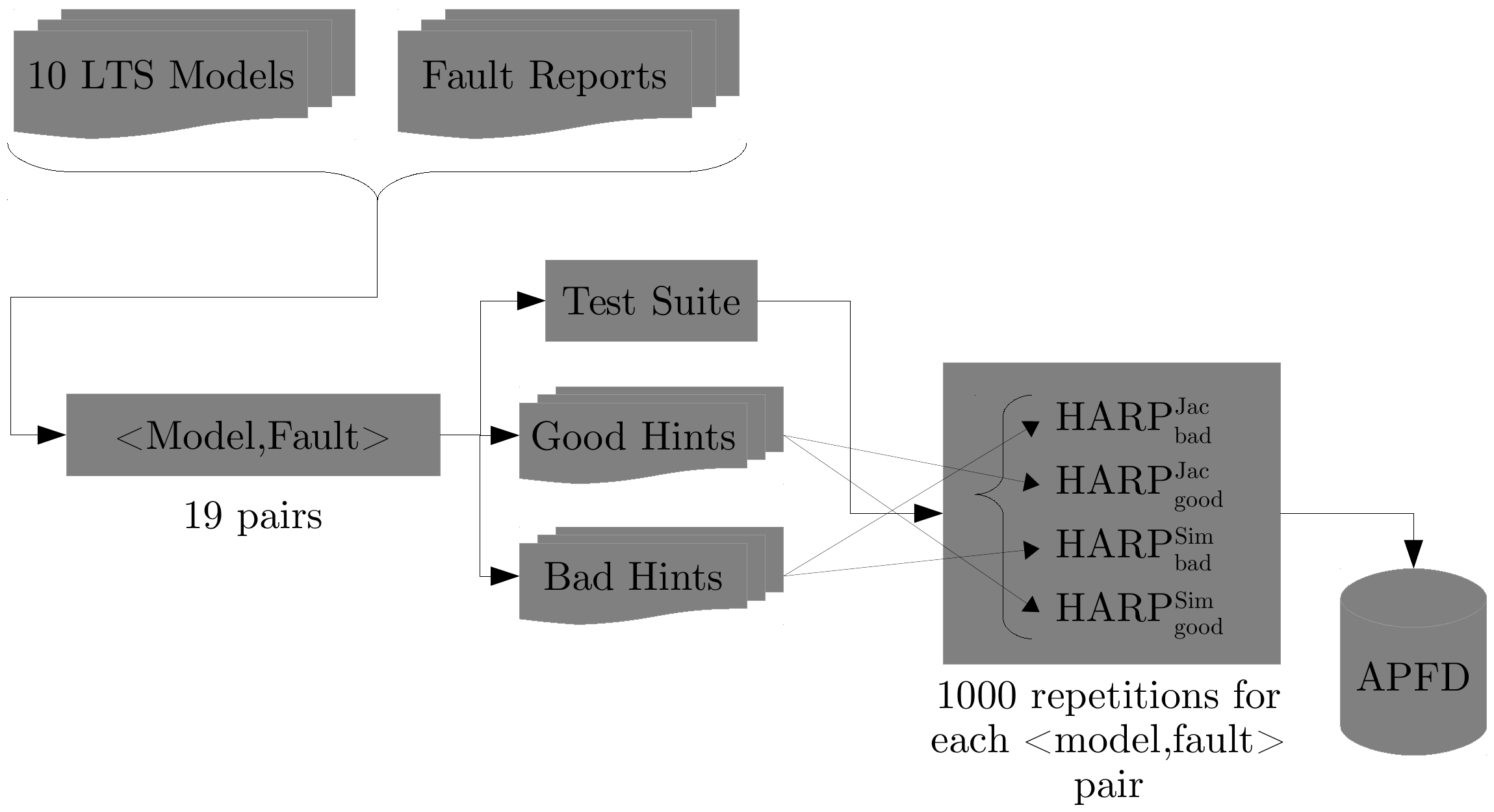}}
\caption{Experiment Design Overview}
\label{fig:Overview}
\end{figure}

Our experimental units are the pair $<$HARP algorithm, resemblance function$>$ and the treatments are both good and bad hints. In order to increase precision and power of statistical analysis, we repeated independently the application of each experimental unit and treatment to every experiment object 1000 times, according guidelines proposed by Arcuri and Briand \cite{ArcuriB2011}. Figure~\ref{fig:Overview} presents the experiment overview.

\subsubsection{Results and Analysis}

\begin{figure}
\centering
\includegraphics[width=0.8\linewidth]{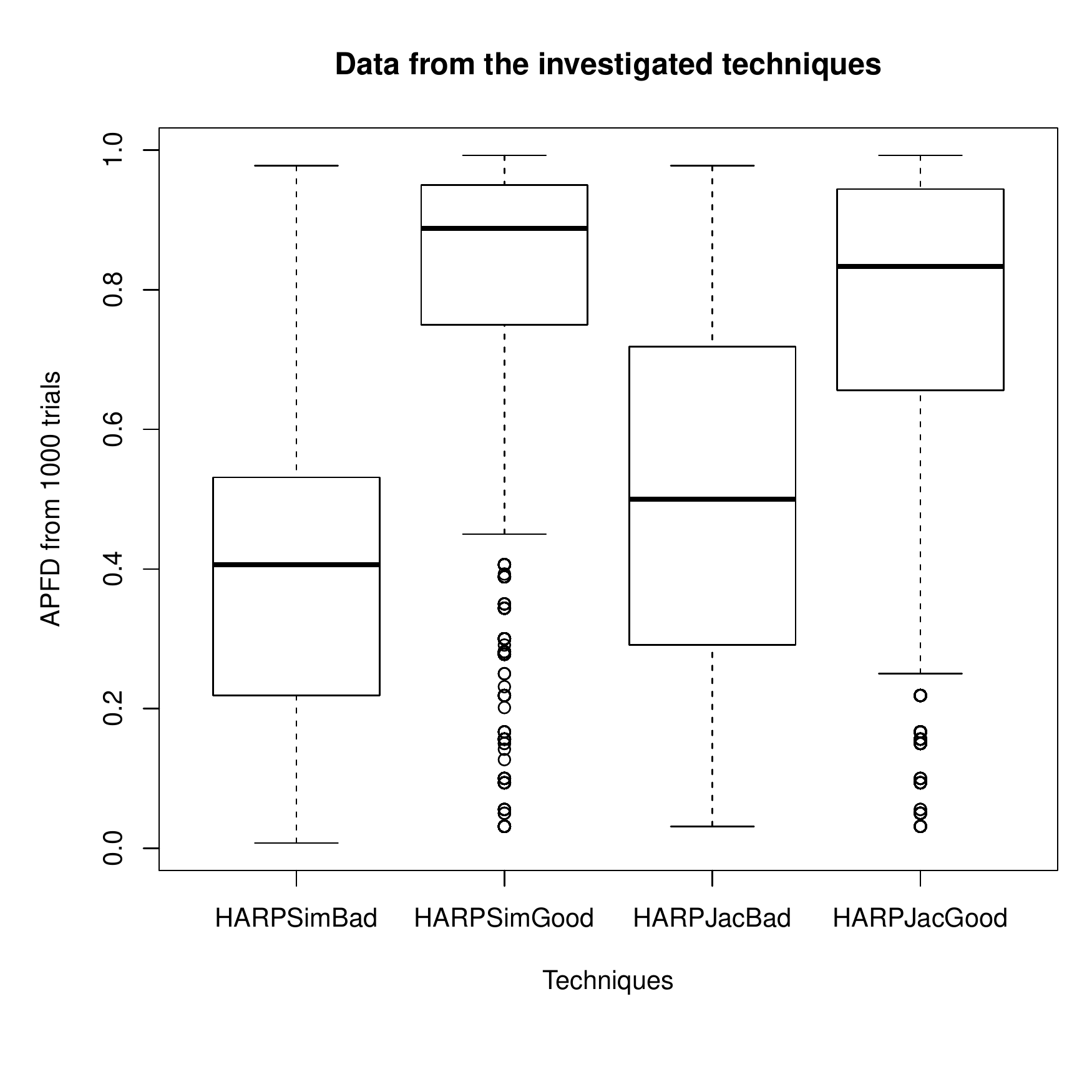}
\caption{Boxplots of the raw data collected in the experiment execution.}
\label{fig:APFDRawData}
\end{figure}

As a first insight about the collected data, we perform a visual analysis (see Figure~\ref{fig:APFDRawData}). From this figure we remark: i) HARP with good hints are more accurate because of a more compact boxplot, however some outliers appear; ii) it is already noticeable the difference between good and bad hints, regardless the applied resemblance function; iii) both resemblance functions appear to have a similar behavior across the levels of the variable hint quality.

Considering both alternatives that receive good hints (\textit{HARPSimGood} and \textit{HARPJacGood}), one is able to notice some outliers, marked as small hollow dots in the respective boxplots. It happens mainly because the HARP's candidate set generation may take many iterations to select a hint related test case. It varies based on the sampled values from the uniform random variable that guides the process. Although very uncommon, the outliers are not mistakes, therefore they must compose the sample to investigate the differences through the effect sizes.

\paragraph{Effect Size Analysis}

To measure the effect sizes and hence properly compare the treatments, we perform a set of parwise A-Tests, which assumes values between 0 and 1, which represent the probability of the result of a direct comparison between two alternatives be right. Table~\ref{tab:rq3effSizes} contains each one of the calculated values. About the statistic value (third column), we can interpret the effect size as \cite{VarghaD00}:

\begin{itemize}
	\item \textit{large}, if the statistic is greater than 0.71 or less than 0.29;
	\item \textit{medium}, if the statistic is greater than 0.64 or less than 0.36;
	\item \textit{small or negligible}, otherwise.
\end{itemize}

\begin{table}
	\centering
	\scriptsize
	\caption{Effect sizes of pairwise comparisons}
	\label{tab:rq3effSizes}
	\begin{tabular}{|c|c|c|c|}
		\hline
		\textbf{Function - Hint} & \textbf{Technique B} & $\mathbf{\hat{A}_{12}}$ & \textbf{Eff. Size} \\ \hline
		Similarity - Bad         & Similarity - Good    & 0.0713                     & Large                \\ \hline
		Similarity - Bad         & Jaccard - Bad        & 0.3628                     & Small                \\ \hline
		Similarity - Bad         & Jaccard - Good       & 0.1271                     & Large                \\ \hline
		Similarity - Good        & Jaccard - Bad        & 0.8484                     & Large                \\ \hline
		Similarity - Good        & Jaccard - Good       & 0.5801                     & Small                \\ \hline
		Jaccard - Bad            & Jaccard - Good       & 0.2237                     & Large                \\ \hline
	\end{tabular}
\end{table}

Addressing \textbf{RQ1}, we evaluate the effect sizes separately between the pairs HARPSimBad | HARPSimGood and HARPJacBad | HARPJacGood. The A statistic value for these comparisons are 0.0713 and 0.2237 respectively, which suggests that the hint's quality has a large effect on HARP. Therefore, HARP performs significantly better when receiving good hints. 

As a further analysis, we observe whether the investigated resemblance functions act differently during HARP operation. To do so, we compare the effect sizes of the pairs HARPSimBad | HARPJacBad and HARPSimGood | HARPJacGood, i.e. change the resemblance function but keeping the level of the variable hint quality. The values for A statistic are 0.3628 and 0.58 respectively. Both values indicate a small effect size, although the first one is close to the medium effect size threshold. Therefore, we do not have enough evidence to affirm that any function is better or more indicated to be used in HARP, in other words both functions express the same notion of resemblance. 

\paragraph{Remarks}

Since we are suggesting to use HARP just with hints pointed by a majority of the team members, allied with the results of the questionnaire reported in Section~\ref{sec:questionnaire}, we believe that situations that a bad hint misguide HARP's operation are minimized. As further analysis, through a simple visual analysis, one can note a smaller interquartile distance when comparing the pairs (HARPSimGood, HARPJacGood) and (HARPSimBad, HARPJacBad) favoring the alternative equipped with the similarity function. Besides it is already focused on MBT, we suggest the investigated similarity function as more indicated to our context.

\subsubsection{Validity Evaluation}

We evaluate the validity of our experiment by discussing its threats and how we deal with them. Concerning the \textbf{internal validity}, the variation of APFD might be also affected by random choices during the techniques execution. In order to deal with this threat, we repeated the executions according to Arcuri and Briand's work \cite{ArcuriB2011}. Besides, the execution order for each technique in every repetition is defined randomly.

About \textbf{construct validity}, since we generate the hints based on prior knowledge about faults and test cases that fail, the process of defining good and bad hints may be biased. Aiming to mitigate this risk, we provide a systematic procedure to do that and we submit both good and bad hints to the same treatments in order to evenly expose them.

Concerning the \textbf{external validity}, even though we use real and industrial systems, a sample with only two systems are not enough to provide proper generalization. Although we were not able to generalize the results due to sample size, we try to be as most realistic as possible by using models that represent the investigated applications, so we believe that similar conclusions would hold for similar systems.

\vspace{-4pt}
\subsection{Questionnaire}
\label{sec:questionnaire}
\vspace{-2pt}

Introducing the idea of getting information from the development team, we perform a questionnaire~\cite{AdamsC08} to collect indications about portions of the system that suffered some occurrence during its development, which are the hints, and investigate whether these hints are related to faults. This is an observational method that helps to assess opinions and attitudes of a sample of interest. In this section, we discuss its methodology and results.

\paragraph{Participants and Investigated Systems}

For this investigation, our population of interest are small development teams that have an interaction with both industry and academy. Therefore, we consider teams from two industrial projects of the Ingenico do Brasil\footnote{www.ingenico.com.br}. SAFF is an information system that manages status reports of embedded devices; and TMA is a system that controls the execution of tests on a series of hardware parts and manages their results. More details regarding these systems, as well as their behaviors, cannot be unveiled due to a non-disclosure agreement (NDA).

\begin{figure}
	\centering
	
	\subfloat[Positions in project]{
		\label{fig:positionInProject}
		\includegraphics[width=.3\textwidth]{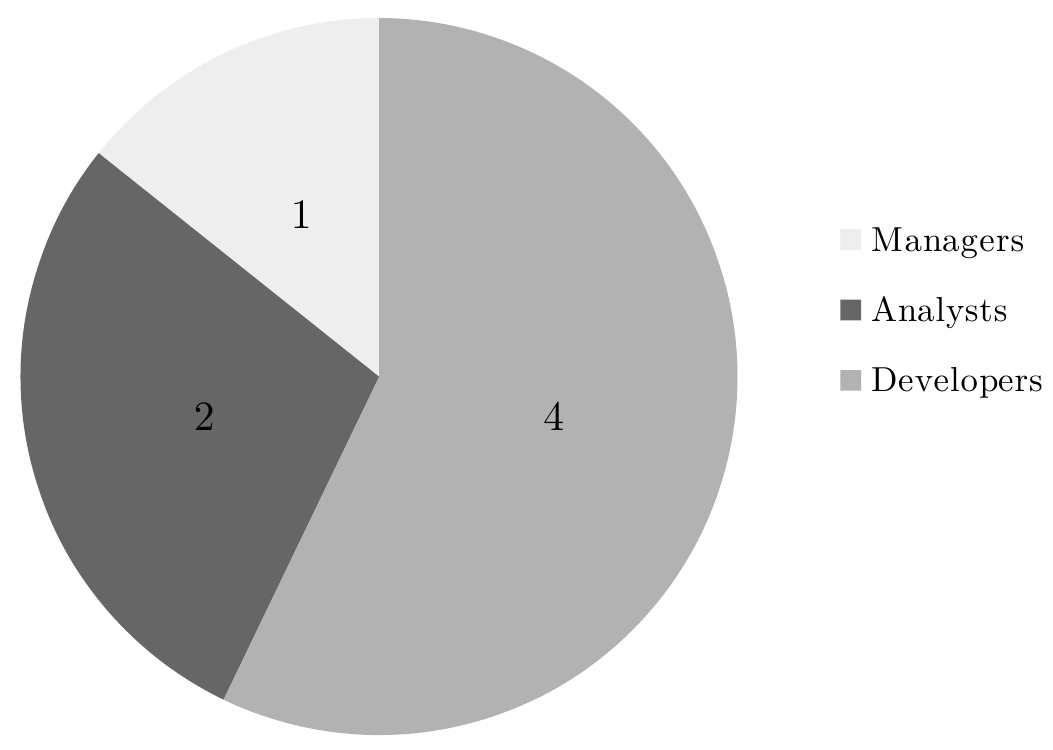}
	}
	\hspace{0.05\textwidth}
	\subfloat[Time in project (months)]{
		\label{fig:timeInProject}
		\includegraphics[width=.3\textwidth]{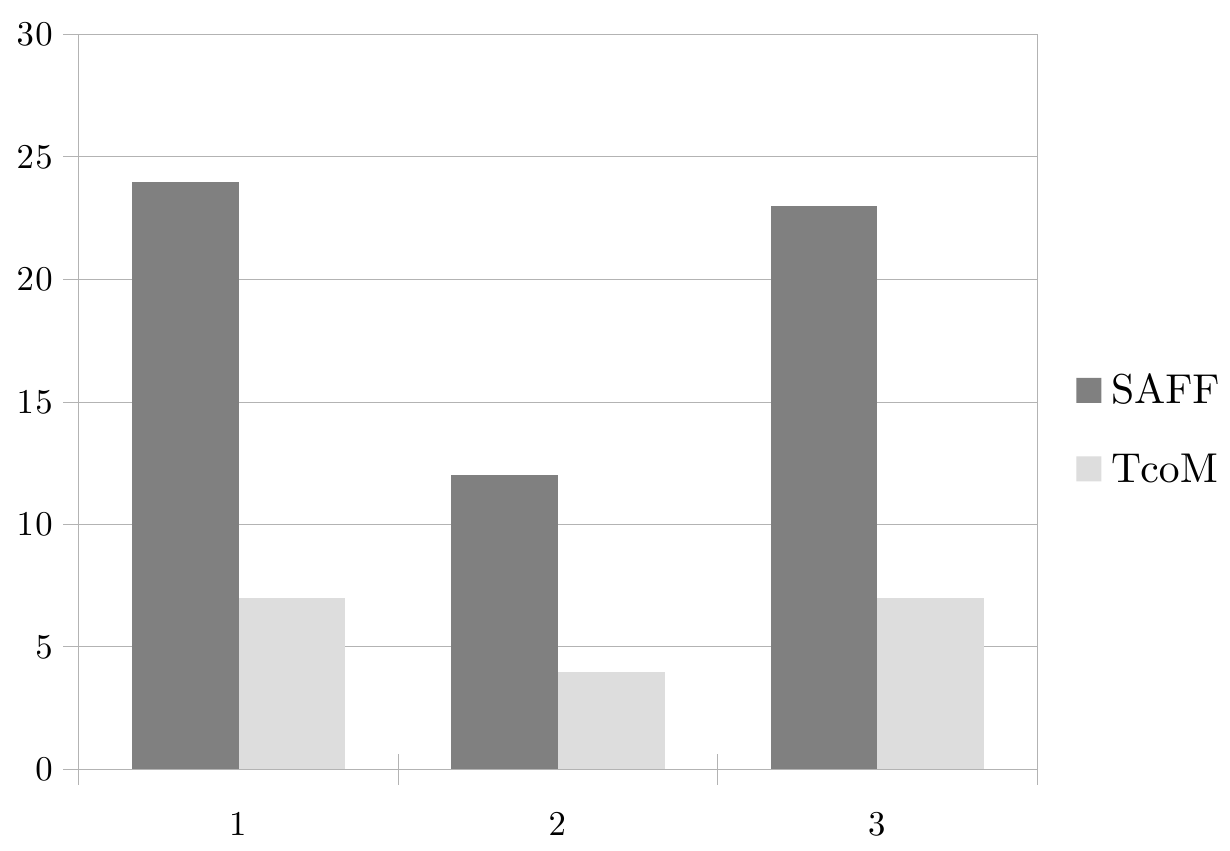}
	}
	\caption{Summary about the questionnaire's participants.}
	\label{fig:participants}
\end{figure}

Since it is hard to find real world teams available for participating on our investigation, and as we do not intend to generalize the results for a larger population, we establish a \textbf{non-probability sampling strategy}~\cite{Kasunic05}. With that sample, we believe that the teams performing in research laboratories in partnership with companies are well represented. Therefore, four members from SAFF and three from TMA compose our sample, as reported in Figure~\ref{fig:positionInProject}. They present different maturity levels, varying from undergraduate students with development experience to professionals. SAFF is an ongoing project and is been under development for the last two years, while TMA in the last seven months. However, one participant of each team started working on the project after it had started, which balances the participants experience in the projects. Figure~\ref{fig:timeInProject} summarizes the time of activity of the participants in their projects. It should be noted that, the participants acted in the development of the investigated use cases using the model-based process described by Jorge et al. \cite{JorgeMNCO14}, in other words the participants designed and/or implemented the use cases considered in this study, therefore they actually faced the reported problems.

\begin{figure}
	\centering
	\includegraphics[width=0.8\linewidth]{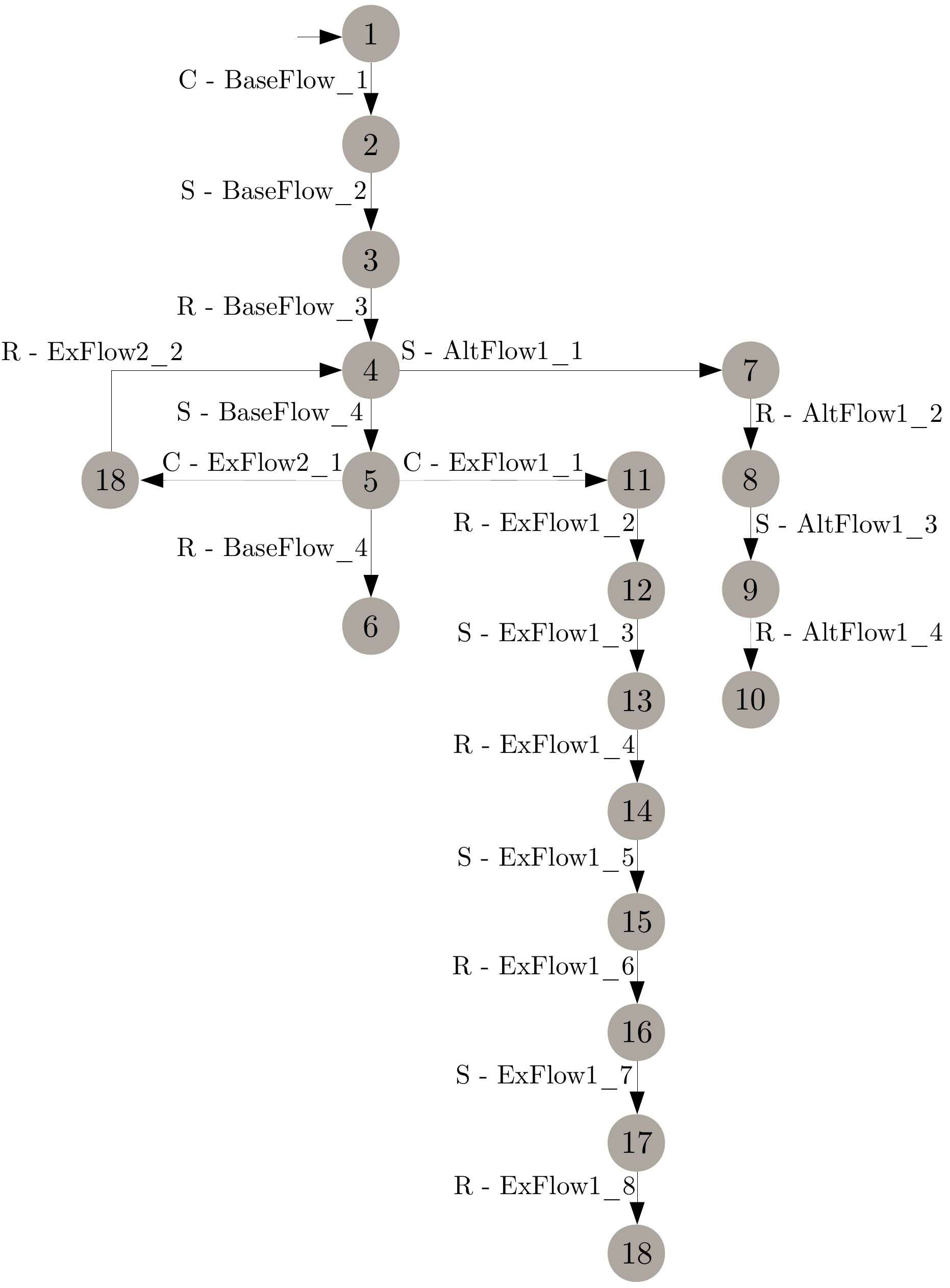}
	\caption{Example of an use case used in the questionnaire, modeled using Labeled Transition System, but with obfuscated labels.}
	\label{fig:useCaseExample}
\end{figure}

To select the target use cases, we check the requirement documents of both systems, which are described using a natural language based use case notation and previous system testing execution results. Among them, we select use cases that have at least one reported fault. 
Among the considered use cases, and aiming to keep a balanced design, we selected two of them for each system. Note that there are seven participants and different amounts of participants from both teams participated. It happened because, whereas for TMA the three participants worked in both use cases, for SAFF one developer worked in one of the investigated use cases and other developer in another, which leads to four participants. To illustrate the dimension of the use cases used here, Figure~\ref{fig:useCaseExample} depicts one of them. It comprises four execution flows - one base flow, two exception, and one alternative - and the exact amount of transitions and states of the original document, but with obfuscated labels in order to hide the actual behavior due to the aforementioned NDA.

\paragraph{Questionnaire}

In order to address \textbf{RQ2} and \textbf{RQ3}, we devise a questionnaire intending to reveal three kinds of information:

\begin{itemize}
	\item \textbf{Experience}: we ask the participant's name, his/her position in the team, and how long the participant had been working in the project. The intention is to evaluate whether the experience of the team really impacts on the ability to give hints;
	
	\item \textbf{Use case critical regions}: we attach to the questionnaire the correspondent use case document and we ask the participant to underline a region in the use case (a single step or an entire flow) that he/she believes that was jeopardized during the development activities before system level testing. For that, we instruct them to solely use his/her experience acquired when implementing the use cases. Moreover, we ask them to provide possible reasons for the choice. We provide a list of reasons: \textit{Inherent complexity} -- when the use case presented aspects that leaded to a complex implementation; \textit{Neglected due to schedule} -- when some other demand had to be satisfied in advance and the schedule became shorter; and \textit{Already present any sort of problem beforehand} -- when the indicated step or flow presented a non-expected behavior in early steps of development and unit testing. Besides, the participants are free to write down any other reason they find adequate. Comparing the answers of this question with the faults and failures reported, we are able to know if it is really possible to gather good hints;
	\item \textbf{How time-consuming and hard was to perform the task}: we also ask to measure the time in minutes they spend to read the attached artifacts and to answer the questionnaire, and whether the participant considers it difficult to read the provided use case document and to answer the related questions. We want to evaluate whether the whole process to obtain hints is feasible.
\end{itemize}

\subsubsection{Results and Analysis}

In order to address \textbf{RQ2}, we resort to the portion of the questionnaire in which the participants underline their hints and explain their reasons. We guide our analysis by observing whether the major part of the participants indicate the same portions with the same reason. A summary of this analysis is available in Table~\ref{tab:answers}, and the bold face cells mark the situations that the participants assign the same hint, i.e. they indicated the same portion of the use case and the same cause.

\begin{table*}
\centering
\footnotesize
\caption{The indications collected from the questionnaires, as well as their reasons. In bold we highlight regions that a majority is achieved.}
\label{tab:answers}
\begin{tabular}{|c|c|c|c|c|}
\hline
\multicolumn{2}{|c|}{}                           & Participant 1                & Participant 2                & Participant 3                \\ \hline
\multirow{4}{*}{SAFF} & \multirow{2}{*}{UC1} & \textbf{Base Flow/Step 4}    & Base Flow/Step 3             & \textbf{Base Flow/Step 4}    \\
                          &                      & \textbf{Inherent Complexity} & Inherent Complexity          & \textbf{Inherent Complexity} \\ \cline{2-5} 
                          & \multirow{2}{*}{UC2} & Exception Flow 1/Step 7      & \textbf{Exception Flow 1}    & \textbf{Exception Flow 1}    \\
                          &                      & Previous Problems            & \textbf{Inherent Complexity} & \textbf{Inherent Complexity} \\ \hline
\multirow{4}{*}{TMA} & \multirow{2}{*}{UC1} & Alternative Flow 4/Step 1    & \textbf{Base Flow/Step 4}    & \textbf{Base Flow/Step 4}    \\
                          &                      & Inherent Complexity          & \textbf{Inherent Complexity} & \textbf{Inherent Complexity} \\ \cline{2-5} 
                          & \multirow{2}{*}{UC2} & --                           & \textbf{Precondition}                 & \textbf{Precondition}                 \\
                          &                      & No Difficulty                & \textbf{No Difficulty}                & \textbf{No Difficulty}                \\ \hline
\end{tabular}
\end{table*}

Considering the Use Case 1 from SAFF (UC1), the participants report an inherent complexity problem hint, but two of them point the same step. Analyzing the fault report related to this use case, we verify that the step pointed by the majority of the participants is directly related to a fault. In turn, for the Use Case 2 (UC2), there are divergences about the reasons of the indications. Two participants report an inherent complexity whereas the other assigned it to problems that already appeared during the development and that is not unveiled on early tests (unit testing). However, when relating it to the actual fault, all participants successfully report the same flow, and it is also related to a fault. Therefore, for this use case, the participants also provide a useful hint.

Now, considering the first use case (UC1) of TMA project, the participants report that an inherent complexity regarding the underlined region could be the source of some problem. Two of them point the same step in the base flow and this step is related to a fault, according to the fault reports. On the other hand, concerning Use Case 2, a different but possible scenario emerges. Every participant wrote freely that they had no difficulty developing this use case, but two of them report the use case's precondition as the most affected region, and the other one do not provide any hint. The fault report contained a fault for this use case, but it is not related to the provided answers. In a conversation out of investigation they discussed that the related precondition is hard to be satisfied, since some network requirements should be met. This is a situation that requires further investigation, since it is hard to measure how a precondition can be related to fault detection.

Considering the use cases where a majority of participants indicate the same portion with the same cause, allied to the fact that every participant worked in the respective use case, in three out of four, they provide a good hint, i.e., a region that actually contained a fault. Therefore, based on our results, we are able to infer that members of development teams, with similar characteristics, may provide good hints about functionalities that they have worked with.

To address \textbf{RQ3}, we analyzed a quantitative measure in minutes collected from the questionnaires and the answer of a question about how hard they think that is to perform the questionnaire. All participants found the task of finding error-prone regions in a use case easy to perform. Moreover, the whole questionnaire was applied on an average of 4.17 minutes when considering the SAFF use cases, and 3.67 for the TMA ones. 

Considering these results, we suggest that the impact of using this approach in other projects tend to be low. Similar questionnaires could be easily used in other industrial projects without much effort. We believe these questionnaires should be handed right between a development and a system testing stage, which is the moment the team still has fresh knowledge regarding what was recently developed. Alternatively, these hints could be collected iteratively during the development phase and then used right before the test case execution to prioritize the system level test cases.

\paragraph{Remarks}

Based on the questionnaires, we can state that when most developers and managers indicate similar hints, these are often good estimators of faults and failures. Thus, we suggest that hints should only be used when pointed by a most of the developers and managers directly involved with the correspondent code, which serves also as a guideline for using HARP. In a tie situation about a particular hint, managers and testers could evaluate and deliberate about the case.

\subsubsection{Validity Evaluation}

In this section, we discuss aspects that may represent some threat to the validity of our study. For instance, when analyzing our results, we must be aware that the participants answered the questionnaire after the use cases were implemented. Therefore, we rely on their memory regarding the system. Moreover, as we intended to investigate developers from projects mixing undergraduates and professionals in cooperation with industry, we were limited to a small and selective sample. 


Due to our limited sampling strategy, we are not able to generalize our results to different contexts. However, since all elements involved in this study report to real projects (e.g., real developers, systems, and use cases), we believe our results are still valid. Nonetheless, our results suggest that similar questionnaires can be used in different contexts.

\vspace{-4pt}
\subsection{Case Study} 
\label{sec:casestudy}
\vspace{-2pt}

After controlling the hint's quality in a controlled environment, in order to address RQ4 we compare the performance of HARP with our baseline, which is the adaptive random prioritization technique proposed by Jiang et al.~\cite{jiang1}. For the comparison, we consider the systems and hints provided in the questionnaire, reported in Section~\ref{sec:questionnaire}. The objective is to evaluate the application of actual hints in industrial systems, observing the differences between \textbf{HARP} and its baseline (\textbf{ARTJac}).

\paragraph{Setup}

We follow the system testing activities on SAFF and TMA systems, which comprised executing manually the \textbf{available test cases} on the applications, which run on small terminals with a proprietary operational system, and \textbf{recording their results}, associating the test cases that failed and a cause described in high level of abstraction. Therefore, we collected these artifacts and used them in our case study. To illustrate the dimensions of the executed test suites, Table~\ref{tab:tsinfo} exposes some relevant testing related metrics.

\begin{table*}
	\centering
	\small
	\caption{Metrics about the test suites investigated in the case study. We measure the shortest and longest test cases in amount of steps, expected results and conditions that the test case describes.}
	\label{tab:tsinfo}
	\begin{tabular}{|c|c|c|c|c|c|c|}
		\hline
		System        & \textbf{\#TC} & \textbf{Shortest TC} & \textbf{Longest TC} & \textbf{\#Faults} & \textbf{\#TC that fail} & \textbf{\#TC filtered by hint} \\ \hline
		\textbf{SAFF} & 60            & 3                          & 89                        & 9                 & 13                      & 6                            \\ \hline
		\textbf{TMA}  & 32            & 7                          & 41                        & 5                 & 8                       & 8                            \\ \hline
	\end{tabular}
\end{table*}

In order to use the provided hints in this case study, we use the manual process of converting them into test purposes as explained in Section~\ref{sec:technique}. However, the amount of hints for each system is different because, there was a case in the TMA system where the participants did not report any problem, therefore we consider just the hint for use case 1.

In this case study, just HARP uses hints to guide its operation, and both techniques take the two aforementioned test suites as input. We need to repeat the executions of these techniques because they both make random choices. Therefore, we run each technique 1000 times independently with the same input, as suggested by Arcuri and Briand \cite{ArcuriB2011}. 

As a measure of fault detection capability, instead of measuring the APFD, which takes into consideration the detection rate of all faults that the test suite is able to unveil, we use the F-Measure \cite{Zhou10}, which considers the ability of detecting the first fault. F-Measure counts the amount of test cases executed before revealing the first fault. A test case reveals a fault when it fails, in other words, when the system produces outputs different from the expected. We are using F-Measure instead of APFD because the participants provided just a single hint for each use case in the questionnaire and when applying this hints in the whole test suite, would be unfair that the other use cases not investigated affect the measurements.

\subsubsection{Results and Analysis}

\begin{figure}
	\centering
	
	\subfloat[Boxplot with the results from SAFF system]{
		\label{fig:saff}
		\includegraphics[width=.45\textwidth]{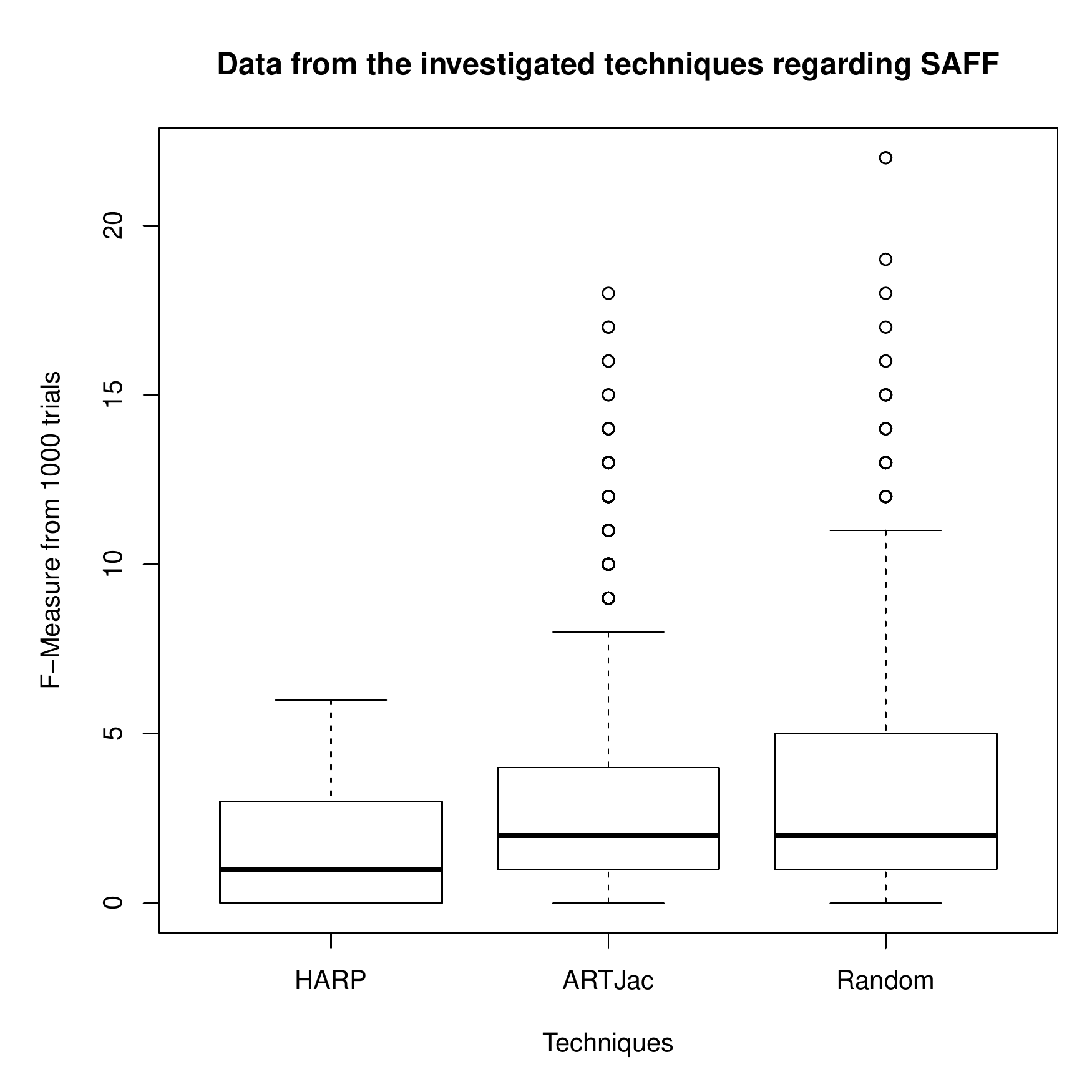}
	}
	\hspace{0.05\textwidth}
	\subfloat[Boxplot with the results from TMA system]{
		\label{fig:tma}
		\includegraphics[width=.45\textwidth]{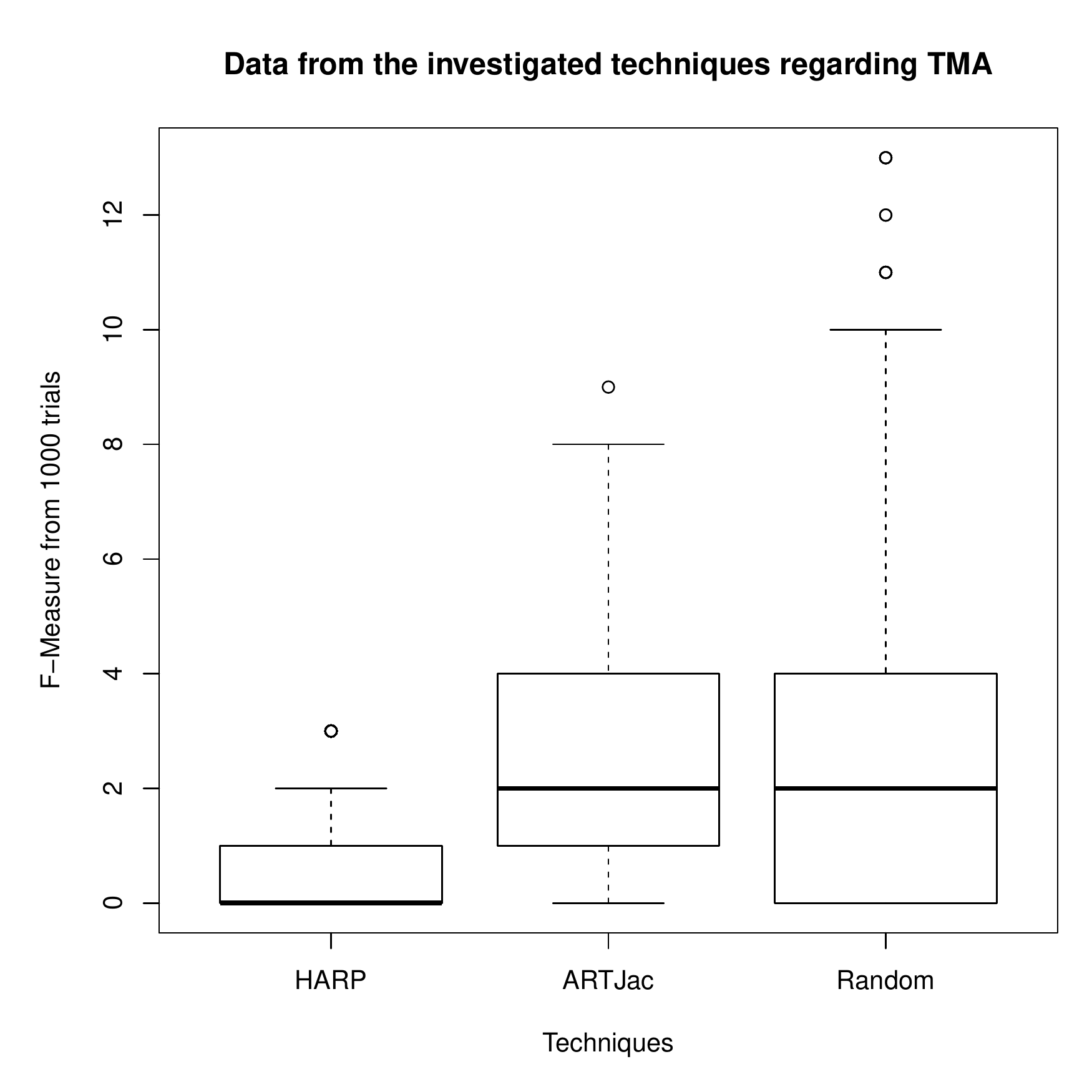}
	}
	\caption{Summary about the collected data.}
	\label{fig:fmeasuredata}
\end{figure}

A visual representation of the collected data is depicted in Figure~\ref{fig:fmeasuredata}. Besides HARP and ARTJac, we added the Random technique as a lower boundary for techniques' performance. A good performance presents a low F-Measure value, which ranges between 0, when the first test case already reveals a fault, to $n-1$, when the last test case reveals a fault. Since both test suites have different sizes, we analyze both of them separately.

Regarding SAFF, based on how spread are the boxplots in Figure~\ref{fig:saff}, HARP presents a more accurate performance than the other two techniques, presenting fewer outliers; however visually it is not possible to conclude which of the techniques present the lowest F-Measure. By calculating the effect size (as performed in Section~\ref{sec:experiment}) of the comparison between HARP and ARTJac we obtain 0.4106 favoring HARP, but with a small effect size. It suggests small gains through the application of HARP.

On the other hand, analyzing visually the boxplots from Figure~\ref{fig:tma}, the improvements either in accuracy or in earlier fault detection are clearer. The effect size measurement of the comparison between them is 0.2442, which means a large effect favoring HARP. Thus, considering these two systems and addressing RQ4, HARP is able to outperform ARTJac when using actual hints.

\paragraph{Remarks}

Even though the small differences between the techniques regarding the median of F-Measure, which are the lines in the center of the boxplots in Figure~\ref{fig:fmeasuredata}, this difference may still provide significant savings in the MBT context and in a context with manual test case execution. For instance, during the development of the systems considered in our evaluation, reports of the test case execution show they took between \textcolor{black}{8-60} minutes, including setup, evaluation, and recording the results using tools. In general, a single test case execution took \textcolor{black}{16-27} minutes and any reduction in the amount of test cases executed before revealing faults is representative, in a process point of view.

\subsubsection{Validity Evaluation}

Here we discuss the validity of this case study by indicating existent threats and how we deal with them. We are not able to draw more general conclusions due to the low amount of systems in our case study, however both systems are real and industrial size. Moreover, the hints were provided by the actual development teams. Although we use two hints from SAFF and one from TMA, we are sure that the hints were given by people that really worked in these use cases. THerefore, to reduce the effect of the other use cases' faults in the ability of reveal faults, we use F-Measure instead of APFD in the case study.

\vspace{-4pt}
\section{Related Work}
\label{sec:relatedwork}
\vspace{-2pt}

In this section, we discuss a series of relevant works related to ours. The TCP problem is more associated with the code-based testing context, as discussed by Catal and Mishra~\cite{CatalM12}. \textbf{HARP} on the other hand is designed to be applied with Model-Based test suites, particularly when test cases are generated from transition systems. There is no consensus about the performance of TCP techniques that deal with new test cases or with the absence of historical data about previous test executions \cite{HenardPHJL16}, which is a motivation for our investigation.

As discussed in Section~\ref{sec:background}, TCP techniques that focus on fault detection aim at estimating a surrogate for this information. Some of the most popular techniques use coverage data: code branches and statements \cite{elbaum3, jiang1}, and models elements \cite{SapnaM09, KKT08}. Korel et al.~\cite{KKT08} propose heuristics for TCP using Extended Finite State Machines - EFSM. These heuristics take into consideration modifications performed during system's evolution, hence the work focus on Regression Testing. As results, the authors suggest that some aspects may not significantly affect the performance of the proposed heuristics. For instance, the amount of times that a modified transition of the EFSM is executed by test cases. \textbf{HARP} focus on the general TCP problem, not relying on history of modifications.

Sapna and Mohanty~\cite{SapnaM09} propose a technique to prioritize test cases generated from UML Activity Diagrams based on a fixed relation of weights of the diagram's elements. Fork/join nodes have more weight than branch/merge nodes, whereas branch/merge nodes have more weight than action nodes. The authors define this relation giving more priority to test cases that represent main scenarios of the system, which cover fewer loops, branches and forks. Another technique suitable for activity diagrams is proposed by Kaur et al.~\cite{KaurBS12}, which also calculates weights, however taking into consideration the amount of incoming and outgoing flows of every model element. Contrary to \textbf{HARP} these techniques take the model structure into account.

Kundu et al. \cite{KunduSSM09} proposes STOOP, a technique that manipulates UML Sequence Diagrams. It generates test cases from input models and prioritizes the generated test cases based on structural elements from sequence diagrams, such as messages and edges. They also propose metrics for three prioritization objectives: code coverage, confidence in reliability, and the composition of both. After calculating the metrics for every test case, according the desired objective, the technique sorts the generated test cases respecting a decreasing order. Contrary to \textbf{HARP} the technique couple the test case generation and test case prioritization processes and focus only on structural elements.

Other studies introduce the idea of collecting data from experts, which are people involved with the development of the system. Stallbaum et al. \cite{StallbaumMP08} propose the augmentation of UML activity diagrams with values related to the probability of specific actions in the diagram contains a fault, as well as values associated with the damage caused whether a fault be revealed in that action, deriving risk scores by multiplying them. This work is very related to our strategy because both acquire information from specialists. However, \textbf{HARP} considers a different prioritization objective, which is fault detection in contrast to risk prevention, and a non-numeric input from the specialists. 

Another source of information presented in literature is a set of pairwise comparisons performed by an expert involving a subset of the provided test cases, which guides the operation of the techniques \cite{TonellaAS2006, YooHTS2009}. The former is focused on risk-based strategies and the latter demands a high interaction with the specialists, because the amount of comparisons that the expert must perform. On the other hand, \textbf{HARP} does not demand a high interaction from experts, as discussed in Section~\ref{sec:questionnaire}.

It is also worth to mention some works related to ARP. Jiang and Chan~\cite{JiangC13}, which have actively contributed in this field, tried to overcome the dependence of code elements by proposing the use of the black-box information to guide the prioritization. Their work provides evidences that this strategy can provide good results in the MBT context, since it uses an abstraction of the source-code representation. Other research that uses the adaptive random strategy, but it is focused on regression testing, is performed by Schwartz and Do \cite{SchwartzD2016}. They propose a technique also based on the adaptive random strategy, but using the Analytic Hierarchy Process (AHP) in order to create importance levels among the tested components, using expert knowledge and data from previous testing executions.

There is a field of investigation related to this research, which is the defect prediction~\cite{Catal11}. Xu et al.~\cite{XuLYAJ16} compare 32 methods to point system features more prone to contain faults. The investigated methods use diverse underlying theories to build their prediction models, such as statistic and probabilities, clustering, and logistic regression. All of them have the same common assumption that historical data about source code metrics, test case coverage, and fault reports. The methods, in turn, divide the data into training and testing sets; their performances are compared through the area under the ROC curve, which relates the rate of true positives and negatives, in a detailed empirical comparison. The authors claim that cluster-based methods can achieve good results whereas retrieving a smaller set of potentially faulty system features. Since we assume that historical data is not available, we are not able to use any of these methods, but we understand their importance as a source of information to prioritize regression test cases. Radjenovi\'{c} et al.~\cite{RadjenovicHTV13} surveyed other metrics used to estimate parts of the system more prone to fail, such as object-oriented metrics, traditional source code metrics, and process metrics. Some of them, mainly the process metrics, also depends on historical information, not available in the context of our research.

To the best of our knowledge, there is no other initiative of applying ARP in the MBT context. In previous work, we conducted a set of studies, as already discussed in Section~\ref{sec:introduction} \cite{OuriquesJSERD} including the two variants, with Jaccard \cite{jiang1} and Manhattan \cite{Zhou10} distance functions. Besides our main results, in the referred context, Jaccard function performed better than the Manhattan. This result provides us evidence that code-based results cannot be generalized to model-based ones, since our results contradict the ones presented by Zhou et al. \cite{ZhouSS12}, which encourages researches in both contexts.
\vspace{-4pt}
\section{Conclusions}
\label{sec:conclusions}
\vspace{-2pt}

Test Case Prioritization (TCP) is an activity dedicated to optimize test case execution in the sense of rescheduling the test cases aiming at achieving some testing goal. In this paper, we propose a TCP technique, named \textbf{HARP}, for MBT system level test suites, in a context that historical information regarding failures/faults may not be available. In order to guide prioritization, our technique uses information derived from the expertise of development teams, named ``hints". We represent these hints as test purposes that work as filters, accepting test cases related to the represented information. HARP is based on ARP, which focus on exploring the input test cases using a distance notion. 

In the first study of our empirical investigation, we control the quality of hints applied to HARP, measuring the fault detection capability through the APFD metric. As results, we provide evidence that, HARP performs significantly better when a good hint is provided, instead of a bad one. In other words, the random choices in HARP do not affect significantly the hints' guidance. Yet, as a secondary analysis, we suggest that different distance and/or similarity functions may capture the same resemblance notion. 

Besides, we applied a questionnaire where we ask participants to indicate steps or flows in use case models representing functionalities that, for example, suffered some unexpected schedule modification or contain an inherent complexity to their developers, and related these indications to actual faults. Our results suggest that the questionnaire is a quick and easy way to collect that kind of information. Moreover, comparing the provided answers to the fault records of the investigated use cases, we observed that the participants were able to provide good hints. In other words, they were able to provide indications that actually points to faults. In addition, we defined guidelines for using these indications as hints in HARP: whether the majority of the team indicates a common aspect of the system, this particular hint should be translated to a test purpose and used in HARP. 

Furthermore, we also perform a case study comparing HARP to its baseline. Results suggest that applying the hints proposed by the development team leads to gains with respect to the ability of detecting faults. In both investigated systems, HARP presented significant improvements against the adaptive random prioritization proposed by Jiang et al. \cite{jiang1}. Even though the measured gains in the first case are not statistically significant, in both situations the practical gains considering the execution time of the test cases are considerable, since in our context the test cases are executed manually.

As future work, we intend to evaluate HARP in a broader variety of systems, against different TCP techniques, and enabling developers and managers to give hints about multiple portions of the system simultaneously, aiming at increasing the validity of our technique. Moreover, we intend to perform other experiment using HARP, but involving different distance/similarity functions. The idea is observing whether these functions represent different resemblance notions and quantifying the effect of these notions on HARP. We also believe that using defect prediction techniques might enable HARP be used in other contexts, such as code-based regression testing.

\bibliographystyle{ieeetr}
\bibliography{bibliography}

\biography{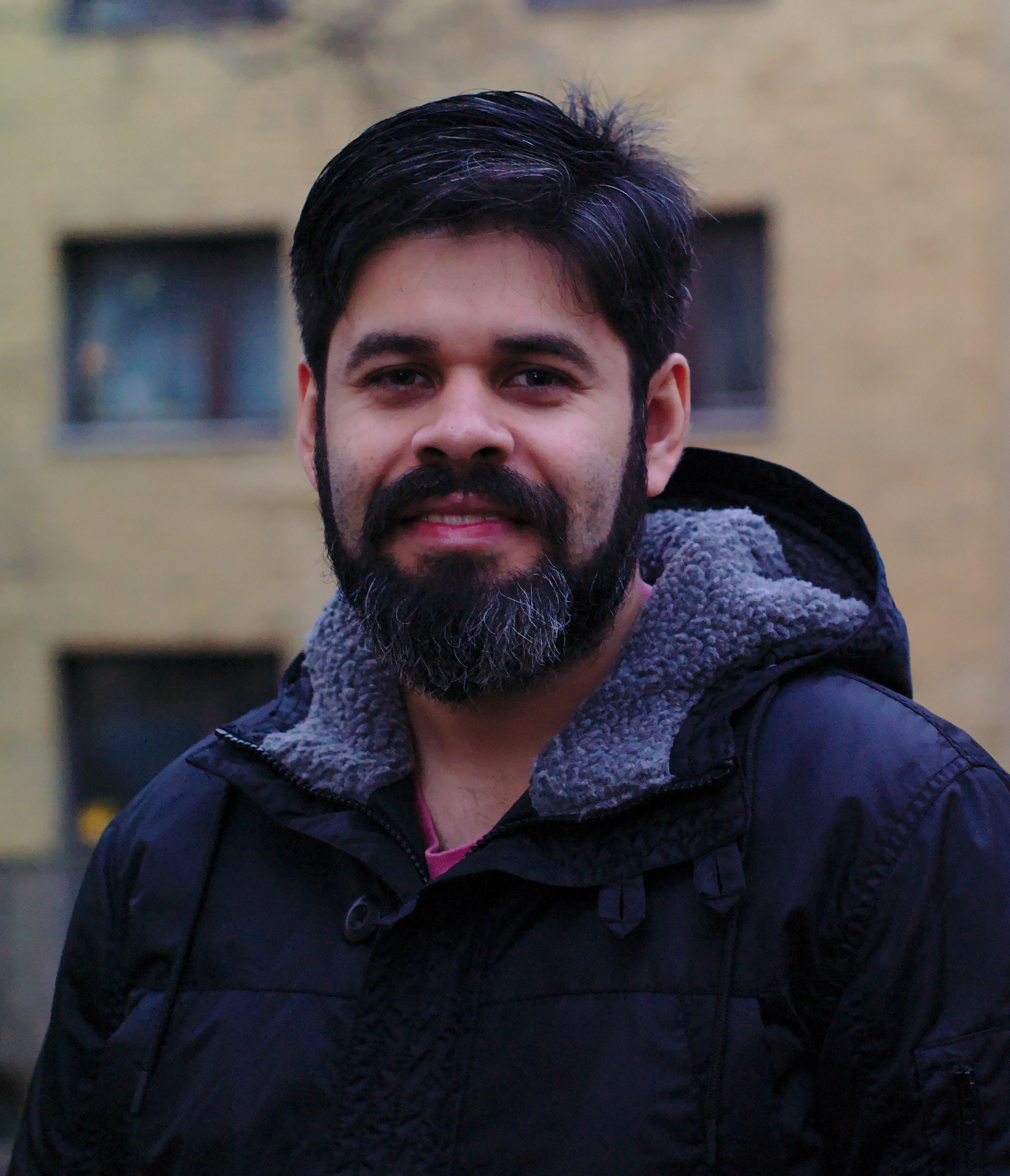}
{\bf Jo\~{a}o Felipe S. Ouriques} is a Ph.D. researcher in the Federal University of Campina Grande (UFCG), since 2012. He holds Master and Bachelor degrees in Computer Science from the Federal University of Campina Grande (UFCG), in 2012 and 2010 respectively. His research interests include software engineering, especially software testing and Mode-Based Testing. In software testing, his focus is on empirical-based investigation of approaches for test case prioritization.

\biography{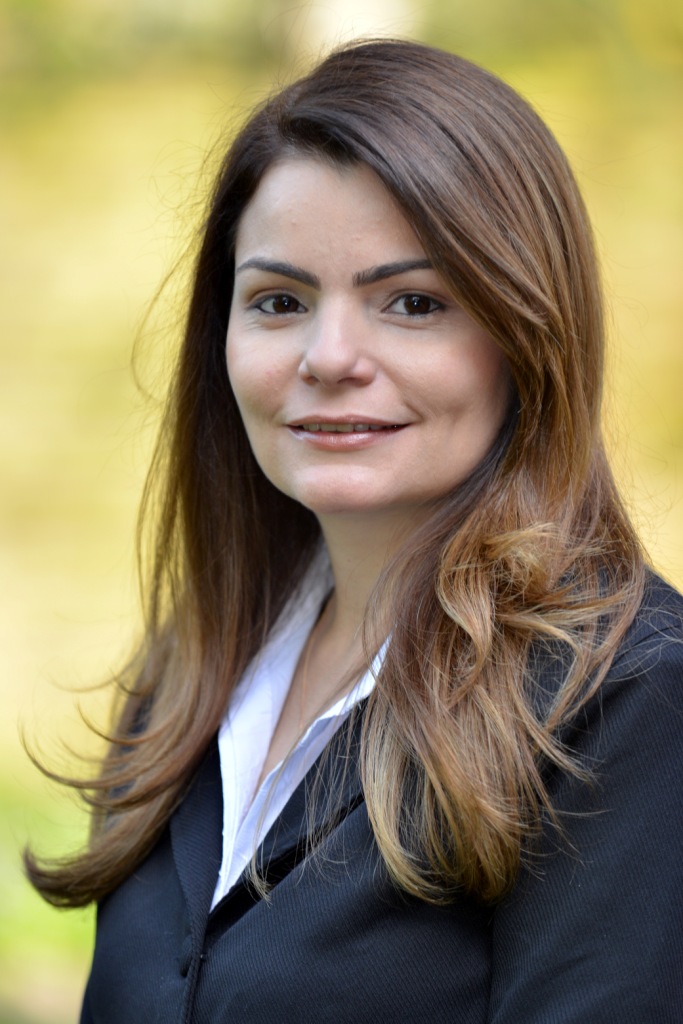}
{\bf Emanuela G. Cartaxo} is an Associate member in the Software Practices Lab at the Federal University of Campina Grande. She received her Doctor Degree in Computer Science from the Federal University of Campina Grande, Brazil, in 2011. Her doctoral research was done at Federal University of Campina Grande in cooperation with Italian National Research Council - CNR (Pisa) and Motorola Brazil Test Center - Research and Development. In 2013, she did a post-doctoral research in the Embedded Systems Quality Assurance Department at Fraunhofer IESE, Germany. Since 2004, she has worked with Software Testing with main focus on Model-Based Testing. Her interests include Software Engineering, especially Software Testing. In Software Testing, the main interest is in Model-Based Testing approaches for test case generation and for controlling the costs of execution.

\biography{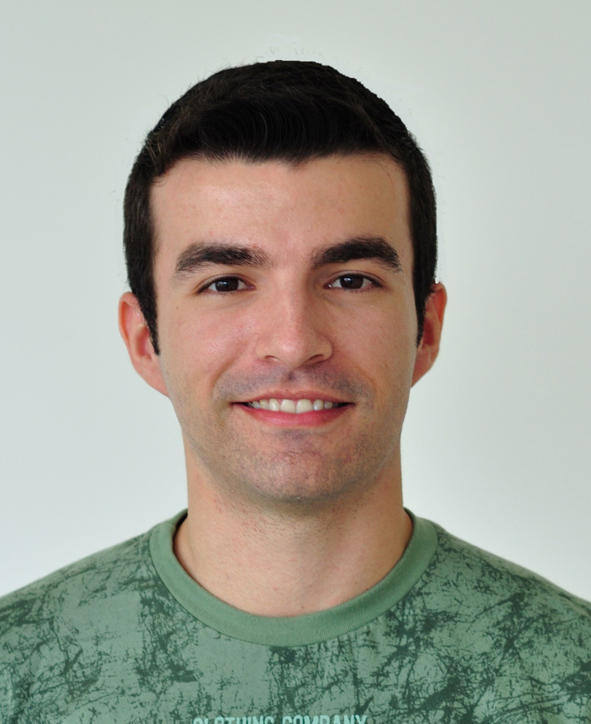}
{\bf Everton L. G. Alves} is a Professor in the Computing and Systems Department at Federal University of Campina Grande (UFCG), Brazil, since 2016. He received his Doctorate degree in Computer Science from the Federal University of Campina Grande, Brazil, in 2015, and his Master and Bachelor Degrees in Computer Science also from the Federal University of Campina Grande, Brazil, in 2011 and 2009, respectively. Between 2013/2014 he worked as visiting researcher at the Department of Electrical and Computer Engineering at the University of Texas at Austin. His main interests include software maintenance, regression testing, model-driven development and testing, real-time systems and integration.

\biography{Machado.eps}
{\bf Patr\'{i}cia D. L. Machado} is a Professor in the Computing and Systems Department at Federal University of Campina Grande (UFCG), Brazil, since 1995. She received her PhD Degree in Computer Science from the University of Edinburgh, UK, in 2001, Master Degree in Computer Science from the Federal University of Pernambuco, Brazil, in 1994 and Bachelor Degree in Computer Science from the Federal University of Paraiba, Brazil, in 1992. Her research interests include software testing, formal methods, mobile computing, component based software development and model-driven development. Since 1998, she has produced a number of contributions in the area of software testing, including research projects, publications, tools, supervising, national/international cooperation and teaching.

\end{multicols}
\end{singlespacing}
\end{document}